\documentclass[11pt]{article}
\usepackage{epsfig,amssymb,a4}
\def\ln{\log}

\def\cconst{C}

\newskip\centreskip
\def\centre#1{\hbox to \hsize{\hskip\centreskip \hbox{#1}\hskip\centreskip}}

\def\color[#1]#2{}

\def\del{\partial}

\def\frac#1#2{\mathinner{#1 \over #2}}

\def\Chi{\hbox{\raise 2pt \hbox{$\chi$}}}
\def\asctilde{\lower 3pt \hbox{\~{}}}

\def\mbf#1{\mathchoice{\hbox{\boldmath $\displaystyle #1$}}
        {\hbox{\boldmath $\textstyle #1$}}{\hbox{\boldmath $\scriptstyle #1$}}
        {\hbox{\boldmath $\scriptscriptstyle #1$}}}

\def\pom{{\mathchoice{\hbox{$\displaystyle\mathbb P$}}
        {\hbox{$\textstyle\mathbb P$}}{\hbox{$\scriptstyle\mathbb P$}}
        {\hbox{$\scriptscriptstyle\mathbb P$}}}}

\def\bfklind{{\hbox{\scriptsize BFKL}}}

\def\Log{\mathop {\csname operator@font\endcsname Log}\nolimits}

\def\putover#1#2{\mathrel{
\setbox0=\hbox{#1}\setbox1=\hbox{\scriptsize #2}
\dimen0=-0.5\wd0 \advance\dimen0 by -0.5\wd1
\dimen1=0.5\wd0 \advance\dimen1 by -0.5\wd1
\hbox{\box0\kern\dimen0%
\vbox to 0pt {\vss\hbox{\raise 0.7em \box1}}%
\kern\dimen1}
}}

\def\putunder#1#2{\mathrel{
\setbox0=\hbox{#1}\setbox1=\hbox{\scriptsize #2}
\dimen0=-0.5\wd0 \advance\dimen0 by -0.5\wd1
\dimen1=0.5\wd0 \advance\dimen1 by -0.5\wd1
\hbox{\box0\kern\dimen0%
\vbox to 0pt {\hbox{\lower 0.7em \box1}\vss}%
\kern\dimen1}
}}

\def\dmath#1#2{
$$\lineskiplimit=1000pt \advance\lineskip by #1\jot 
\mathsurround=0pt \tabskip=0pt plus 1000pt
\everycr{\noalign{\penalty\interdisplaylinepenalty}}
\halign to \displaywidth{
\hfil$\displaystyle{##}$\tabskip=0pt&%
\hfil $\displaystyle{{}##{}}$\hfil &%
$\displaystyle{##}$\hfil \tabskip=0pt plus 1000pt minus 1000pt&%
\refstepcounter{equation}\label{##}\llap{(\theequation)}\tabskip=0pt\cr
\noalign{\ifdim \prevdepth>-1000pt \vskip -#1\jot\fi}
#2\crcr}$$}

\def\crd{\crcr\noalign{\unpenalty
\multiply\interdisplaylinepenalty by 2 \penalty \interdisplaylinepenalty}}

\def\crl#1{\crcr\noalign{\unpenalty\penalty 10000
\nointerlineskip \vbox to 0pt {
\dimen0=\lineskip \vskip \dimen0 minus 1000pt \hbox to \displaywidth{%
\hfil \refstepcounter{equation}\label{#1}(\theequation)}
\vskip 0pt minus 1000pt} \penalty 10000}}

\def\Ez#1{{E^{(\nu_{#1},0)}}}
\def\eEz#1#2#3{{\left(\frac{|\rho_{#1#2}|}{|\rho_{#1#3}||\rho_{#2#3}|}\right)%
^{1+2i\nu_{#3}}}}
\def\cEz#1#2#3{{\left(\frac{|\rho_{#1#2}|}{|\rho_{#1#3}||\rho_{#2#3}|}\right)%
^{1-2i\nu_{#3}}}}

\def\sumabc{\sum_{a\leftrightarrow b\leftrightarrow c}}
\def\crossrat#1#2#3#4#5#6#7#8{%
{\frac{\rho_{#1#2}\rho_{#3#4}}{\rho_{#5#6}\rho_{#7#8}}}}
\def\icrossrat#1#2#3#4#5#6#7#8{%
{(\rho_{#1#2}\rho_{#3#4})/(\rho_{#5#6}\rho_{#7#8})}}
\def\psithree#1#2#3#4#5#6#7#8{%
{\left(\frac{|\rho_{#1#2}||\rho_{#3#4}|}{|\rho_{#5#6}|}\right)^{1#72i\nu_{#8}}
}}
\def\onetworat#1#2#3#4#5#6#7#8{%
{\left(\frac{|\rho_{12}|}{|\rho_{1#8}||\rho_{2#8}|}
\frac{|\rho_{#1#2}||\rho_{#3#4}|}{|\rho_{#5#6}|}\right)^{1#72i\nu_{#8}}
}}
\def\onethreerat#1#2#3#4#5#6#7#8{%
{\left(\frac{|\rho_{13}|}{|\rho_{1#8}||\rho_{3#8}|}
\frac{|\rho_{#1#2}||\rho_{#3#4}|}{|\rho_{#5#6}|}\right)^{1#72i\nu_{#8}}
}}
\def\twothreerat#1#2#3#4#5#6#7#8{%
{\left(\frac{|\rho_{23}|}{|\rho_{2#8}||\rho_{3#8}|}
\frac{|\rho_{#1#2}||\rho_{#3#4}|}{|\rho_{#5#6}|}\right)^{1#72i\nu_{#8}}
}}

\def\be{\begin{equation}}
\def\ee{\end{equation}}
\def\bea{\begin{eqnarray}}
\def\eea{\end{eqnarray}}
\def\nn{\nonumber}
\def\hbar{\bar{h}}

\def\kf{{\mbf k}}

\newdimen\picraise
\newcommand\picbox[1]
{
 \setbox0=\hbox{\input{#1}}
  \picraise=-0.5\ht0
  \advance\picraise by 0.5\dp0
  \advance\picraise by 3pt     
  \hbox{\raise\picraise \box0}
}

\begin{document}
\begin{titlepage}
\begin{flushright}
HD-THEP-03-33\\
hep-ph/0308056
\\
\end{flushright}
\vfill
\begin{center}
\boldmath
{\LARGE{\bf How Pomerons Meet in Coloured Glass}}
\unboldmath
\end{center}
\vspace{1.2cm}
\begin{center}
{\bf \Large
Carlo Ewerz $^a$, Volker Schatz $^b$
}
\end{center}
\vspace{.2cm}
\begin{center}
{\sl
Institut f\"ur Theoretische Physik, Universit\"at Heidelberg\\
Philosophenweg 16, D-69120 Heidelberg, Germany}
\end{center}                                                                 
\vfill
\begin{abstract}
\noindent
We compute the perturbative one-to-three Pomeron vertex 
in the colour glass condensate using the extended generalized leading 
logarithmic approximation in high energy QCD. 
The vertex is shown to be a conformal four-point 
function in two-dimensional impact parameter space. 
Our result can be used to compare different theoretical 
descriptions of the colour glass condensate. \\

\noindent
\vfill
\end{abstract}
\vspace{5em}
\hrule width 5.cm
\vspace*{.5em}
{\small \noindent 
$^a$ email: C.Ewerz@thphys.uni-heidelberg.de \\
$^b$ email: V.Schatz@thphys.uni-heidelberg.de 
}
\end{titlepage}

\section{Introduction}
\label{sec:intro}

In hadronic collisions at high energy the partons inside the 
colliding hadrons form a very dense system which is known 
as the colour glass condensate. The colour glass condensate is 
expected to have very interesting properties which can be 
studied experimentally in deep inelastic lepton-hadron scattering 
as well as in hadron-hadron collisions. One aspect 
of particular interest is the question whether the parton 
densities inside the colliding hadrons saturate at sufficiently 
high energies. If saturation indeed takes place it can in turn 
tame the growth of hadronic cross sections with the 
centre-of-mass energy $\sqrt{s}$. It is a very challenging 
theoretical problem to determine the relevant scales in the 
energy and in the momentum transfer at which saturation effects 
set in and to predict the resulting behaviour of the cross sections. 

A variety of different approaches has been developed in order 
to find answers to these questions. The approach which has been 
most popular recently and from which the name colour glass condensate 
originates is the one initiated by McLerran and Venugopalan 
\cite{McLerran:1993ka,McLerran:1993ni,McLerran:1994vd} 
which was further developed for example in 
\cite{Jalilian-Marian:1996xn,Iancu:2000hn,Ferreiro:2001qy}, 
for a review and further references see \cite{Iancu:2003xm}. 
A related approach based on the operator expansion method applied to 
Wilson line operators was used in 
\cite{Balitsky:1995ub}-\cite{Balitsky:2001mr}. The relation between 
these two approaches was established in \cite{Blaizot:2002xy}. 
The latter approach leads to a hierarchy of equations which can be 
drastically simplified in the large-$N_c$ limit. In this limit one obtains 
the so-called Balitsky-Kovchegov equation which has also been 
obtained in \cite{Kovchegov:1999yj} using the colour dipole approach 
to high energy scattering \cite{Mueller:1993rr}-\cite{Chen:1995pa}. 
In the colour dipole approach one studies the evolution of a small 
colour dipole due to gluon emissions. In the large-$N_c$ limit the 
emission of a gluon can be interpreted as a splitting of the dipole into 
two dipoles, and the interaction of two such systems after evolution 
is described by the basic dipole-dipole scattering via gluon exchange. 

Another approach to the physics of the colour glass condensate is 
the classic one based on the perturbative resummation of large 
logarithms of the centre-of-mass energy $\sqrt{s}$ in high energy 
collisions. This resummation program 
was performed in \cite{Kuraev:fs,Balitsky:ic}, resulting in the 
celebrated Balitsky-Fadin-Kuraev-Lipatov 
(BFKL) equation which resums all terms of the 
form $(\alpha_s \log s)^n$ in the perturbative series. 
This approximation scheme is called the leading logarithmic 
approximation (LLA). On the level of the cross section, i.\,e.\ 
the squared amplitude, the BFKL equation describes the exchange 
of two interacting reggeised gluons in the $t$-channel, called 
the BFKL Pomeron. It is known that at higher energies 
exchanges of more than two gluons in the $t$-channel 
eventually become important, and one has to include them in 
order to find an appropriate description of the high energy limit. 
This is done in the generalized leading logarithmic approximation 
(GLLA). There are two forms of this approximation. The first 
one considers only exchanges in which the number of reggeised 
gluons remains constant during the $t$-channel evolution. 
The corresponding resummation of the leading logarithmic 
terms for a given number of gluons is encoded in the 
Bartels-Kwieci{\'n}ski-Prasza{\l}owicz (BKP) equations 
\cite{Bartels:1980pe,Kwiecinski:1980wb}. That form of the 
GLLA can be considered as the quantum mechanics of 
states of $n$ gluons exchanged in the $t$-channel. 
One expects to obtain a scattering amplitude satisfying 
the unitarity bound for the total cross section 
when all contributions with arbitrary $n$ are 
included\footnote{Therefor the $n$-gluon amplitudes of both 
versions of the GLLA are often called unitarity corrections.}. 
In order to obtain amplitudes satisfying unitarity also in 
all subchannels one has to include also contributions in which 
the number of reggeised gluons is permitted to vary during 
the $t$-channel evolution, giving rise to the second version of 
the GLLA which we call extended GLLA (EGLLA) 
\cite{Bartels:1980pe,Bartels:1978fc,Bartels:unp}. 
That step turns the quantum mechanical 
problem of states with a fixed number of gluons into a quantum 
field theory of reggeised gluons in which states with different 
numbers of gluons are coupled to each other. 
The corresponding amplitudes are described by a tower of coupled 
integral equations generalizing the BKP equations. These 
equations constitute the basis of the EGLLA. 
Their structure has been investigated in a series of papers, 
see for example \cite{Bartels:1992ym}-\cite{Volkerthesis}. 
A remarkable result found in these studies is the conformal 
invariance of the amplitudes of the EGLLA in two-dimensional 
impact parameter space. There are in fact strong indications that 
the whole set of amplitudes can be formulated as a conformal 
field theory. 

The approaches mentioned above are to a large extent based 
on the application of perturbative concepts. The use of perturbation 
theory is justified in high energy scattering processes 
which are dominated by a single large momentum scale. 
It would clearly be a great success if one could obtain a good 
description of the high energy limit of QCD 
in a perturbative framework. The practical applicability 
of such a perturbative picture would remain restricted to certain 
scattering processes dominated by a large momentum scale. 
Nevertheless, it would be very valuable to see how for example the 
structure of Regge theory can emerge from perturbative QCD. 
It should be emphasized, however, that the wealth of hadronic scattering 
processes at high energies is dominated by low momenta. A satisfactory 
description of high energy scattering therefore eventually requires 
to include nonperturbative effects. The hope is that one can explore 
the transition region to low momenta starting from the perturbative 
framework. A recent example for the use of this method is 
\cite{Ferreiro:2002kv} where it was found that 
the validity of the Froissart bound on the high energy behaviour 
of total cross sections, $\sigma_T \le \mbox{const}\cdot \log s$, appears 
to be closely related to confinement effects. In the long-term perspective 
one would hope that it is also possible to find relations between 
the perturbative approaches and generically nonperturbative approaches 
to high energy scattering like for example the implementation 
\cite{Kramer:tr}-\cite{Dosch:1994ym} of the stochastic vacuum model of 
\cite{Dosch:1987sk,Dosch:ha,Simonov:1987rn} in the nonperturbative 
framework of \cite{Nachtmann:1991ua}, or even 
Regge theory (see for example \cite{Donnachie:en}), although this 
is extremely challenging. 

The different perturbatively motivated approaches have 
their respective advantages. The approach of 
\cite{McLerran:1993ka}-\cite{Iancu:2003xm} is for example 
particularly well suited for estimating the momentum scales at which 
saturation effects set in whereas in the EGLLA the conformal invariance 
of the amplitudes and the phenomenon of reggeisation become 
particularly transparent. 
It would obviously be very desirable to understand the exact 
relations between these different 
approaches to the problem of high energy scattering. 
Interestingly, the BFKL equation is reproduced by all 
approaches mentioned above in a suitable limit corresponding 
to sufficiently low parton densities. In particular, that limit 
corresponds to a situation in which nonlinear effects in the 
evolution of the system can be neglected. 
Another important quantity which plays an important role 
in all approaches to the colour glass condensate is the 
perturbative one-to-two Pomeron vertex 
which was first derived in the EGLLA
\cite{Bartels:1994jj,Lotter:1996vk}. Its value for the leading 
Pomeron states was computed in \cite{Korchemsky:1997fy} 
and was found to agree with the value found in the dipole model 
of high energy scattering \cite{Bialas:1997xp,Bialas:1997ig}. 
The very same vertex occurs also in the other approaches 
and is a crucial ingredient for example in the 
Balitsky-Kovchegov equation. In order to find characteristic 
differences between the different approaches one should 
therefore go beyond the approximation in which only Pomerons 
and their simplest vertex occurs. One possibility is to look 
at the exchange of negative charge parity quantum numbers 
associated with the Odderon, for a review see \cite{Ewerz:2003xi}. 
The Odderon has only been identified and studied in 
the two versions of the GLLA, and most recently also in 
the dipole picture \cite{Kovchegov:2003dm}. The other promising 
possibility is to study vertices with more than three Pomerons 
and here in particular the one-to-three Pomeron vertex. 
This vertex has so far only been considered in the dipole model 
and in the approach due to Balitsky, and there are apparently 
contradicting findings about its existence in the literature. 
Based on a generalization of the one-to-two Pomeron vertex 
in the dipole model an explicit form of the one-to-three Pomeron 
vertex is conjectured in \cite{Peschanski:1997yx}, whereas 
in \cite{Braun:1997nu} it is argued that such a vertex is absent 
in the dipole picture at least in the leading approximation. 
In the approach due to Balitsky, on the other hand, a one-to-three 
Pomeron transition was identified in \cite{Balitsky:2001mr}. 
In the present paper we derive the one-to-three Pomeron 
vertex in the colour glass condensate using the framework of the 
EGLLA and hence establish the existence of such a vertex 
in this approach. We obtain the one-to-three Pomeron vertex 
by projecting the two-to-six reggeised gluon vertex calculated in 
\cite{Bartels:1999aw} onto Pomeron eigenfunctions. 

The paper is organised as follows. In section \ref{sec:eglla} 
we recall some important properties of the EGLLA 
and describe in particular the effective two-to-six reggeon 
vertex. We also discuss the main results for the one-to-two 
Pomeron vertex and the available results and and conjectures 
concerning vertices with more Pomerons. 
We describe how the effective reggeon vertices of the 
EGLLA can be used to calculate multi-Pomeron 
vertices, in particular how the one-to-three Pomeron vertex 
is obtained from the two-to-six reggeon vertex. 
In section \ref{sec:Wp3p} we perform the explicit 
calculation of the one-to-three Pomeron vertex. 
Collecting the different contributions obtained in that section we 
give our final result for the one-to-three Pomeron 
vertex and discuss its implications 
in section \ref{sumandconcl}. Some useful formulae needed 
for the calculation are derived in the appendix. 

\section{Extended GLLA, effective reggeised gluon vertices and 
Pomeron vertices}
\label{sec:eglla}

\subsection{Extended GLLA}
\label{subsec:eglla}

In the EGLLA one considers exchanges in the $t$-channel 
in which the number of reggeised gluons can fluctuate during 
the $t$-channel evolution. 
The objects of interest in the EGLLA are amplitudes 
describing the production of $n$ reggeised gluons in the 
$t$-channel. For any given $n$ one then resums all diagrams 
of the perturbative series which contain the maximal number 
of logarithms of the energy for that $n$. So far 
the EGLLA has been studied explicitly only for the case 
that the system of reggeised gluons is coupled to a virtual photon 
impact factor. 
It is expected that due to high energy factorization the results obtained 
in that special case are universal. In particular, the interaction 
vertices between states with different numbers of reggeised gluons 
have a universal meaning independent of the impact factor. 

The amplitudes $D_n^{a_1\dots a_n}(\kf_1, \dots, \kf_n)$ describe 
the production of $n$ reggeised gluons in the $t$-channel carrying 
transverse momenta $\kf_i$ and colour labels $a_i$. 
The lowest order term of these amplitudes 
is given by the virtual photon impact factor consisting of a quark loop 
to which the $n$ gluons are coupled. But there are also terms in 
which less than $n$ gluons are coupled to the quark loop and 
there are transitions to more gluons during the $t$-channel 
evolution. Technically this means that the amplitudes $D_n$ obey 
a tower of coupled integral equations in which the equation for a given 
$n$ involves all amplitudes $D_m$ with $m<n$. The first of these 
integral equations (i.\,e.\ the one for $D_2$) is identical to 
the BFKL equation. In the higher equations new transition kernels 
occur which are generalizations of the BFKL kernel. They have 
been derived in \cite{Bartels:unp}. 
Various aspects of the system of the integral equations have been 
studied in \cite{Bartels:1992ym}-\cite{Volkerthesis}, for a detailed 
description and a systematic approach to solving the equations 
see \cite{Bartels:1999aw}. 

The structure of the solutions is such that they involve only 
states with fixed even numbers of gluons which are coupled to each 
other by effective transition vertices that can be computed explicitly. 
The reggeisation of the gluon is responsible for the 
fact that in the solutions of the integral equations for the amplitudes 
$D_n$ the states with fixed odd numbers of gluons do not occur. 
The simplest example for reggeisation in the amplitudes is the 
three-gluon amplitude which can be shown to be a superposition 
of two-gluon amplitudes of the form 
\be
\label{d3reggeizes}
D^{abc}_3 (\kf_1,\kf_2,\kf_3) = g f_{abc} 
\left[
D_2(\kf_1+\kf_2,\kf_3) - D_2(\kf_1+\kf_3,\kf_2) 
+ D_2(\kf_1,\kf_2+\kf_3) 
\right]
\,.
\ee
As a consequence, an actual 
three--gluon state in the $t$-channel does not occur. 
In each of the three terms 
in this expression the momenta of two gluons enter only 
as a sum. They can be regarded as 
forming a `more composite' reggeised gluon that occurs 
in the amplitude $D_2$. 
Reggeisation occurs also in all higher amplitudes $D_n$, 
all of which contain a contribution which can be decomposed 
into two-gluon amplitudes in a way similar to 
(\ref{d3reggeizes}). 
In the higher amplitudes also more than two gluons can form 
a more composite reggeised gluon. The details of this process are 
discussed in \cite{Ewerz:2001fb}. 

The three-gluon amplitude can be written completely in terms 
of two-gluon amplitudes. In the higher amplitudes with $n\ge 4$ 
gluons additional contributions occur. The structure of the four-gluon 
amplitude can be written in symbolic form as 
\cite{Bartels:1993ih,Bartels:1994jj}: 
\begin{equation}
\label{d4solutionpics}
D_4 = \sum 
\begin{picture}(0,0)%
\includegraphics{solutiond41.pstex}%
\end{picture}%
\setlength{\unitlength}{3947sp}%
\begingroup\makeatletter\ifx\SetFigFont\undefined%
\gdef\SetFigFont#1#2#3#4#5{%
  \reset@font\fontsize{#1}{#2pt}%
  \fontfamily{#3}\fontseries{#4}\fontshape{#5}%
  \selectfont}%
\fi\endgroup%
\begin{picture}(865,966)(2373,-2016)
\end{picture}
 + \begin{picture}(0,0)%
\includegraphics{solu42.pstex}%
\end{picture}%
\setlength{\unitlength}{4144sp}%
\begingroup\makeatletter\ifx\SetFigFont\undefined%
\gdef\SetFigFont#1#2#3#4#5{%
  \reset@font\fontsize{#1}{#2pt}%
  \fontfamily{#3}\fontseries{#4}\fontshape{#5}%
  \selectfont}%
\fi\endgroup%
\begin{picture}(907,1398)(3875,-2389)
\put(4681,-1951){\makebox(0,0)[lb]{\smash{\SetFigFont{10}{12.0}{\rmdefault}{\mddefault}{\updefault}$V_{2 \to 4}$}}}
\end{picture}
 
\hspace*{.5cm}\,.
\end{equation}
Here the first term is again the reggeising part consisting of a 
superposition of two-gluon amplitudes in a way similar to 
(\ref{d3reggeizes}). In the second term the $t$-channel evolution 
starts with a two--gluon state that is coupled to a full four-gluon 
state via a new effective 2-to-4 gluon transition vertex 
$V_{2 \rightarrow 4}$ the explicit form of which was found in 
\cite{Bartels:1993ih,Bartels:1994jj}. 
As can be seen from (\ref{d4solutionpics}) 
the four-gluon state does not couple directly to the quark box diagram 
of the photon impact factor. 

The structure emerging here is that of a quantum field theory in which 
states with different numbers of gluons are coupled to each other 
via effective transition vertices like $V_{2 \rightarrow 4}$. 
This structure has been shown to persist also to higher amplitudes in 
\cite{Bartels:1999aw}. There the amplitudes with up to six gluons 
were studied. It turns out that the five-gluon amplitude can 
be written completely in terms of two-gluon amplitudes and 
four-gluon amplitudes containing the vertex $V_{2 \rightarrow 4}$. 
The mechanism of reggeisation is universal and it is expected 
that all amplitudes $D_n$ with an odd number $n$ of gluons 
are superpositions of lower amplitudes with even numbers of gluons. 
The six-gluon amplitude $D_6$ was shown to consist of three 
parts. First there are two reggeising parts, one of them (called 
$D_6^R$ in \cite{Bartels:1999aw}) is a superposition of two-gluon 
amplitudes. In this term the six gluons arrange themselves 
in all possible ways in two groups which can then be regarded 
as composite gluon states forming the two reggeised gluons 
of the two-gluon amplitude. The second part of the six-gluon 
amplitude (called $D_6^{IR}$ in \cite{Bartels:1999aw}) is 
a superposition of four-gluon states. In these the six gluons 
arrange themselves in four groups, and the resulting 
amplitude is identical to a superposition of four-gluon amplitudes. 
The four gluons couple to the photons in a way similar 
to the last term in (\ref{d4solutionpics}), namely via a 
two-gluon state. 
The coupling of the two-gluon to the four-gluon state 
happens via 
the transition vertices $I,J$ and $L$ of \cite{Bartels:1999aw} 
which can in turn be related to the two-to-four gluon 
vertex $V_{2\to 4}$ in (\ref{d4solutionpics}). 
These are the two reggeising parts of the six-gluon amplitude. 
Finally, there is a part in which a two--gluon state is coupled to a 
six-gluon state via a new effective transition vertex $V_{2 \to 6}$ 
which has been computed explicitly in \cite{Bartels:1999aw}. 
This vertex will be the starting point for our calculation of the 
one-to-three Pomeron vertex. We will therefore describe it 
in more detail in the next section. 
Closing this section we point out that 
the effective transition vertices have the remarkable property  
of being conformally invariant in two-dimensional impact parameter 
space \cite{Bartels:1995kf,Ewerz:2001uq}. 

\subsection{The $2\to6$ reggeised gluon vertex}
\label{sec:V2to6}

Let us now review the properties of the two-to-six 
reggeised gluon vertex occurring in the effective field 
theory structure found for the $n$-gluon amplitudes $D_n$. 
Here we concentrate on the properties 
which are essential for the calculation of the one-to-three 
Pomeron vertex, for the derivation and further properties 
of the two-to-six gluon vertex we refer the reader to 
\cite{Bartels:1999aw,Ewerz:2001uq}. 

The two-to-six vertex couples a two-gluon state to a six-gluon 
state. The two-gluon state is given by a BFKL Pomeron amplitude. 
In \cite{Bartels:1999aw} only the case was considered in which 
that Pomeron amplitude is coupled to a quark loop. The corresponding 
Pomeron amplitude is denoted by $D_2$. Further below we will instead 
use a Pomeron amplitude $\Phi_2$ without the quark loop 
and then assume it to be an eigenfunction of the BFKL kernel. 
The two-to-six vertex is defined as an integral operator acting 
on the two-gluon state $D_2$. It carries six colour labels for the 
outgoing gluons, and the incoming gluons are in a colour 
singlet state. Further the vertex is a function of the transverse 
momenta of the incoming and outgoing gluons. When the vertex acts on 
the two-gluon state $D_2$ one integrates over the transverse momenta 
of the two incoming gluons. We will use a shorthand notation in 
which the six momentum arguments are replaced by the indices 
of the respective gluons, 
\be
(V_{2\to6}^{a_1a_2a_3a_4a_5a_6}D_2)
(\kf_1,\kf_2,\kf_3,\kf_4,\kf_5,\kf_6)=
(V_{2\to6}^{a_1a_2a_3a_4a_5a_6}D_2)(1,2,3,4,5,6)
\,,
\ee
i.\,e.\ 1 stands 
for $\kf_1$, 2 for $\kf_2$ and so on.  A string of several indices, for instance 123, 
means the sum of the momenta, $\kf_1+\kf_2+ \kf_3$.  An empty argument, which 
will be denoted by ``$-$'', stands for zero momentum.  This notation has the 
advantage that it can be easily transformed into a configuration-space 
expression.  In configuration space, an index stands for the 
corresponding coordinate.  A string of indices means that the coordinates are 
identified via two-dimensional delta functions by which the expression has 
to be multiplied.  An empty argument ``$-$'' translates into an integral over 
the corresponding coordinate argument in position space. 

The dependence of the vertex on the colour labels of the six outgoing 
gluons is given by 
\dmath2{
(V_{2\to6}^{a_1a_2a_3a_4a_5a_6}D_2)(1,2,3,4,5,6)=&&
        d_{a_1a_2a_3}d_{a_4a_5a_6}(WD_2)(1,2,3;4,5,6) \crd
        &+& d_{a_1a_2a_4}d_{a_3a_5a_6}(WD_2)(1,2,4;3,5,6)+\ldots
\,.
&eq:V2to6}%
It contains only colour tensors which are products of two 
symmetric structure constants $d_{abc}$ of $su(3)$.  
The sum runs over all ten partitions of the six momenta and 
the corresponding colour indices into two groups of three.  

The function $(WD_2)$
is a convolution of a function $W$ and a Pomeron amplitude $D_2$.  
As we have already pointed out the vertex occurs only in situations in which 
it acts on a BFKL Pomeron state, 
and we will therefore consider only the combination $(WD_2)$ 
rather than the integral operator $W$ alone.
Note that the function $(WD_2)$ is the same in all terms in 
(\ref{eq:V2to6}), only the momentum arguments are permutated 
in the different terms. 
As shown in \cite{Ewerz:2001uq} the function $(WD_2)$ can be 
expressed in terms of a function $G$, 
\dmath2{
(WD_2)(1,2,3;4,5,6)&=&
\frac{g^4}8 \sum_{M\in\mathcal P(\{1,\ldots,6\})}
(-1)^{\#M}\;G(123\backslash M,M,456\backslash M)
\,.
&eq:WD2}%
This sum runs over all subsets $M$ of the set of indices. The notation
``123$\backslash M$'' means the indices 1, 2 and 3 except those contained
in~$M$.  $\#M$~is the number of indices contained in~$M$.  This sum and the
symmetry of the function~$G$ under exchange of its first and third argument
cause the function $(WD_2)$ to be symmetric under all permutations of its first
three and its last three arguments, and under exchange of these two groups of
arguments.  Together with the sum in~(\ref{eq:V2to6}), this makes the $2\to6$
vertex symmetric under arbitrary permutations of the six outgoing 
gluons, i.\,e.\ under simultaneous 
permutations of the colour labels and momenta of the gluons. 

The function $G$ had been introduced in \cite{Bartels:1994jj} and was 
further analysed in \cite{Braun:1997nu,Volkerthesis}. It depends on 
three momentum arguments and has the following form: 
\dmath{0.5}{
G(1,2,3)&=&
\frac{g^2}2\big[2\,c(123)-2\,b(12,3)-2\,b(23,1)+2\,a(2,1,3)
\cr
&&\qquad
{}+t(12,3)+t(23,1)-s(2,1,3)-s(2,3,1)\big]\,,
&eq:G_symb}%
with component functions $a$, $b$, $c$, $s$, and $t$ which will be 
defined below. It can be easily seen already from~(\ref{eq:G_symb}) 
that $G$ is symmetric under the exchange of its first and third argument.  
That symmetry does not apply to the second argument.  
However, there is a different interesting property related to that argument. 
When the second momentum argument is set to zero, the $a$ and $s$ functions 
vanish and $G$ essentially reduces to the well-known BFKL 
kernel applied to the amplitude $D_2$, 
\bea
\label{greducemomentum}
G(1,-,3)&=& 
\frac{1}{N_c} \kf_1^2 \kf_3^2 (K_\bfklind\otimes D_2)(1,3) \,,
\nn \\
&=& \frac{g^2}{2} [ 2 c(13) - 2 b(1,3) - 2 b(3,1) +t(1,3) + t(3,1)]             
\,.
\eea
This is becomes clear directly from the explicit form of the 
component functions to which we turn momentarily, and 
which also defines the convolution in this equation. 
Note that we understand the BFKL kernel to contain a 
delta function identifying the sums of the two incoming and 
the two outgoing momenta. 
Further, it can be proven that $G$ vanishes when its first or last momentum 
argument vanishes, 
\dmath{0.5}{
G(1,2,-)&=&G(-,2,3)=0 \,.
&eq:G-23}%
This is because then the component functions either vanish
or turn into a different component function with fewer arguments (see below) 
and cancel among each other.

Let us now take a closer look at the component functions $a$, $b$, $c$, $s$ 
and~$t$, of which $G$ is composed. They were originally defined in 
momentum space \cite{Bartels:1999aw} via 
\dmath{2.5}{
a(\mbf k_1,\mbf k_2,\mbf k_3)&=&\int\frac{d^2\mbf l}{(2\pi)^3}
        \frac{\mbf k_1^2}{(\mbf l-\mbf k_2)^2(\mbf l-\mbf k_1-\mbf k_2)^2}
        D_2(\mbf l,\mbf k_1+\mbf k_2+\mbf k_3-\mbf l)
&eq:abcst_m_a\cr
b(\mbf k_1,\mbf k_2)&=&\int\frac{d^2\mbf l}{(2\pi)^3}
        \frac{\mbf k_1^2}{\mbf l^2(\mbf l-\mbf k_1)^2}
        D_2(\mbf l,\mbf k_1+\mbf k_2-\mbf l)
&eq:abcst_m_b\cr
c(\mbf k)&=&\int\frac{d^2\mbf l}{(2\pi)^3}
        \frac{\mbf k^2}{\mbf l^2(\mbf l-\mbf k)^2}D_2(\mbf l,\mbf k-\mbf l)
&eq:abcst_m_c\cr
s(\mbf k_1,\mbf k_2,\mbf k_3)&=&\int\frac{d^2\mbf l}{(2\pi)^3}
        \frac{\mbf k_1^2}{\mbf l^2(\mbf l-\mbf k_1)^2}
        D_2(\mbf k_1+\mbf k_2,\mbf k_3)
&eq:abcst_m_s\cr
t(\mbf k_1,\mbf k_2)&=&\int\frac{d^2\mbf l}{(2\pi)^3}
        \frac{\mbf k_1^2}{\mbf l^2(\mbf l-\mbf k_1)^2}
        D_2(\mbf k_1,\mbf k_2) \,.
&eq:abcst_m_t}%
These functions are not infrared safe separately, but 
the infrared divergences cancel in the combination in which 
they occur in the function $G$. 

There are some relations between the functions $a$, $b$ and~$c$ and between $s$
and $t$ and the so-called trajectory function~$\beta$, namely 
\dmath2{
b(\mbf k_1,\mbf k_2)&=&a(\mbf k_1,0,\mbf k_2)=a(\mbf k_1,\mbf k_2,0)
&eq:abcstrel_m_b\cr
c(\mbf k)&=&b(\mbf k,0)=a(\mbf k,0,0)
&eq:abcstrel_m_c\cr
s(\mbf k_1,\mbf k_2,\mbf k_3)&=&
        -\frac2{N_cg^2}\beta(\mbf k_1)D_2(\mbf k_1+\mbf k_2,\mbf k_3)
&eq:abcstrel_m_s\cr
t(\mbf k_1,\mbf k_2)&=&s(\mbf k_1,0,\mbf k_2)=
        -\frac2{N_cg^2}\beta(\mbf k_1)D_2(\mbf k_1,\mbf k_2)
&eq:abcstrel_m_t}%
The trajectory function~$\beta$ is defined in momentum space as 
\begin{equation}
\label{eq:beta_m}
\beta(\mbf k)=-\frac{N_cg^2}2\int\frac{d^2\mbf l}{(2\pi)^3}
\frac{\mbf k^2}{\mbf l^2(\mbf l-\mbf k)^2}\,, 
\end{equation}
and the function $\alpha(\kf) = 1 + \beta(\kf)$ is known as the 
Regge trajectory of the gluon. 

It is convenient to perform the calculation of the one-to-three 
Pomeron vertex in impact parameter space where one can make use 
of the properties of the function $G$ and of the component functions 
under conformal transformations. 
These properties have first been studied in \cite{Braun:1997nu} 
where it was argued that $G$ is invariant under conformal 
transformations in impact parameter space, i.\,e.\ under 
M\"obius transformations of the coordinate arguments of $G$. 
This property was used in \cite{Ewerz:2001uq} to prove the 
conformal invariance of the two-to-six gluon vertex $V_{2 \to 6}$. 
For a more rigorous discussion of the conformal invariance of 
$G$ and of the behaviour of the regularised component functions 
under conformal transformations see \cite{Volkerthesis}. 

We therefore also need the impact parameter (or position space) 
representation of the component functions. The momentum space 
and positions space representations are related to each other 
via Fourier transformation, for the function $G$ for example 
we have
\be
G(\kf_1,\kf_2,\kf_3) = 
\int \prod_{i=1}^3 \left[ d^2\rho_i \,e^{-i \kf_i \mbf{\rho}_i} \right] 
G(\mbf{\rho}_1,\mbf{\rho}_2,\mbf{\rho}_3)
\,.
\ee
In two-dimensional impact parameter space we use complex 
coordinates, $\rho = \rho_x + i \rho_y$. 
We use the same symbol for the momentum space representation 
and the position space representation of a function, and again 
we often make use of the shorthand notation in which an index now stands 
for the corresponding coordinate. 
Let us now give the expressions for the component functions in 
complex coordinates \cite{Volkerthesis}. For reasons to be explained 
in the next section we here use the notation 
$D_2(\rho_1,\rho_2) = \Delta_1 \Delta_2 \Phi_2 (\rho_1,\rho_2)$. 
Then we have 
{\def\mbf{} 
\dmath{2}{
a(\mbf\rho_1,\mbf\rho_2,\mbf\rho_3)&=&
-\frac1{(2\pi)^3}\Delta_1
[\ln2+\psi(1)-\ln m-\ln|\mbf\rho_{12}|]\cdot{}\crd
&&\qquad\qquad{}\cdot
[\ln2+\psi(1)-\ln m-\ln|\mbf\rho_{13}|]
\Delta_2 \Delta_3 \Phi_2(\mbf\rho_2,\mbf\rho_3)
&eq:abcst_c_a\cr
b(\mbf\rho_1,\mbf\rho_2)&=&\frac1{(2\pi)^2}\Delta_1
[\ln2+\psi(1)-\ln m-\ln|\mbf\rho_{12}|]\,
\Delta_2\,\Phi_2(\mbf\rho_1,\mbf\rho_2)
&eq:abcst_c_b\cr
c(\mbf\rho)&=&-\frac1{2\pi}\Delta\,\Phi_2(\mbf\rho,\mbf\rho)
&eq:abcst_c_c%
\cr
s(\mbf\rho_1,\mbf\rho_2,\mbf\rho_3)&=&
\frac2{(2\pi)^3}
\left(2\pi\,\delta^2(\mbf\rho_{12})\,[\ln2+\psi(1)-\ln m-\ln |\mbf\rho_{12}|]
-\frac1{|\mbf\rho_{12}|^2}\right)\cdot{}\crd
&&\qquad\qquad\qquad{}\cdot
\Delta_2\,\Delta_3\,\Phi_2(\mbf\rho_2,\mbf\rho_3)
&eq:abcst_c_s\cr
t(\mbf\rho_1,\mbf\rho_2)&=&
\frac2{(2\pi)^3}\int d^2\mbf\rho_0
\left(2\pi\,\delta^2(\mbf\rho_{10})\,[\ln2+\psi(1)-\ln m-\ln |\mbf\rho_{10}|]
-\frac1{|\mbf\rho_{10}|^2}\right)\cdot{}\crd
&&\qquad\qquad\qquad\qquad{}\cdot
\Delta_0\,\Delta_2\,\Phi_2(\mbf\rho_0,\mbf\rho_2) \,,
&eq:abcst_c_t}}%
where we have used the definition 
\be
 \rho_{ij} = \rho_i - \rho_j \,.
\ee
These functions are regularised with a gluon mass~$m$, and the limit $m\to0$ is
implied in all expressions above. The expressions in square brackets are 
identical to the expansion of $K_0$ Bessel functions for
small arguments which originate from the configuration-space gluon propagator, 
\begin{equation}
K_0(z)=\ln 2 + \psi(1) - \ln z + \mathcal O(z^2) \,.
\label{eq:K0expand}
\end{equation}
In the following, we will for brevity often choose to write $K_0$ instead 
of the expansion. 

In position space the relations between the five component functions and the 
trajectory function $\beta$ have the following form: 
{\def\mbf{} 
\dmath{1.5}{
b(\mbf\rho_1,\mbf\rho_2)&=&
\int d^2\mbf\rho'\,a(\mbf\rho_1,\mbf\rho',\mbf\rho_2)
=\int d^2\mbf\rho'\,a(\mbf\rho_1,\mbf\rho_2,\mbf\rho')
&eq:abcstrel_c_b\cr
c(\mbf\rho)&=&\int d^2\mbf\rho'\,b(\mbf\rho,\mbf\rho')
=\int d^2\mbf\rho'\,d^2\mbf\rho''\,a(\mbf\rho,\mbf\rho',\mbf\rho'')
&eq:abcstrel_c_c\cr
s(\mbf\rho_1,\mbf\rho_2,\mbf\rho_3)&=&
-\frac2{N_cg^2}\beta(\mbf\rho_{12})\,\Delta_2\,\Delta_3\,
\Phi_2(\mbf\rho_2,\mbf\rho_3)
&eq:abcstrel_c_s\cr
t(\mbf\rho_1,\mbf\rho_2)&=&
\int d^2\mbf\rho_0\;s(\mbf\rho_1,\mbf\rho_0,\mbf\rho_2)=
-\frac{2}{N_cg^2} \int d^2\mbf\rho_0\;
\beta(\mbf\rho_{10})\,\Delta_0\,\Delta_2\,\Phi_2(\mbf\rho_0,\mbf\rho_2) \,.
&eq:abcstrel_c_t}}%
In addition we have the equivalent of the relation (\ref{greducemomentum}) 
in position space. Here the function $G$ reduces to a 
double Laplacian of the well-known BFKL kernel multiplied 
by a constant when the second momentum argument is set to zero, 
\dmath2{
G(1,-,3)&=&
\frac1{N_c}\Delta_1\,\Delta_3\, (K_\bfklind\otimes D_2)(1,3) \,.
&eq:G_BFKL}%

Let us finally note that the representation~(\ref{eq:WD2}) for the two-to-six 
vertex is easily generalised to other (even) numbers of reggeised gluons. 
In fact the $2\to4$ reggeised gluon vertex can be
written as a sum over functions $(VD_2)$ which are defined in terms 
of the function $G$ in analogy 
to~(\ref{eq:WD2})~\cite{Bartels:1994jj}.  In that case the sum runs over all 
(three) partitions of four indices into two groups of two.  It is conjectured that there
exist higher $2\to2n$ reggeised gluon vertices which are also new elements of
the effective field theory of unitarity corrections and which are of analogous
form \cite{Ewerz:2001uq}. 

\subsection{Multi-Pomeron vertices}
\label{sec:pvertices}

We will now use the $2 \to 2n$ reggeised gluon vertices of the 
EGLLA in order to define $1 \to n$ Pomeron vertices. We will give 
the definition explicitly only for $n=2,3$, but assuming the existence 
of the higher $2 \to 2n$ reggeised gluon vertices it is straightforward 
to extend the definition to arbitrary $n$. Due to the conformal 
invariance of the amplitudes of the EGLLA in impact parameter 
space it is most convenient to consider also the multi-Pomeron 
vertices in impact parameter space. 

The vertices of the effective field theory for the $n$-gluon amplitudes 
in the EGLLA couple states with different numbers of reggeised 
gluons to each other. The vertex $V_{2 \to 4}$ for example couples 
a two-gluon state to a general four-gluon state. By construction the 
two-gluon state is a BFKL Pomeron state which is coupled to a 
virtual photon impact factor. The vertex is separated 
in rapidity from that quark loop by Pomeron evolution. Due to high energy 
factorization the vertex is hence independent of the photon impact 
factor. In the following we will therefore use a pure BFKL Pomeron 
amplitude $\Phi_2$ instead of the amplitude $D_2$ 
containing the photon impact factor. More precisely, one usually 
chooses to replace (removing two propagators) 
\be
\label{replaced2phi2}
D_2(\rho_1,\rho_2) \longrightarrow \Delta_1 \Delta_2 \Phi_2 (\rho_1,\rho_2) 
\,,
\ee
and we will follow this convention also for the $1 \to 3$ Pomeron 
vertex further below. 
One then obtains the one-to-two Pomeron vertex from the
$2 \to 4$ reggeised gluon vertex $V_{2 \to 4}$ by choosing the 
four-gluon state to be a product of two BFKL Pomeron states 
\cite{Bartels:1994jj,Lotter:1996vk}. 
In this way one projects two pairs of outgoing gluons onto BFKL 
Pomeron states. 

Due to the conformal invariance of the BFKL kernel 
the BFKL Pomeron can be expanded in conformal partial 
waves \cite{Lipatov:1985uk}. It is therefore sufficient to 
consider the one-to-two Pomeron vertex for these 
conformal partial waves which are given by eigenfunctions 
of the BFKL kernel. The eigenfunctions are given by 
\begin{equation}
\label{eq:Enun_app}
E^{(\nu,n)}(\rho_{1},\rho_{2})=
\left(\frac{\rho_{12}}{\rho_{1a}\rho_{2a}}\right)^h
\left(\frac{\rho_{12}^*}{\rho_{1a}^*\rho_{2a}^*}\right)^{\bar{h}} \,.
\end{equation}
According to the representation theory of the group $SL(2,{\mathbb C})$ of 
conformal transformations in two dimensions the eigenfunctions 
are parametrised by the conformal weight 
\be
h = \frac{1+n}{2} + i \nu 
\,,
\ee
where $n \in {\mathbb Z}$ is the integer conformal spin, and $\nu$ takes on 
arbitrary real values. In (\ref{eq:Enun_app}) we have used 
$\bar{h} = 1 - h^*$. 
The eigenfunctions $E{(\nu,n)}$ of the BFKL kernel depend on an 
additional parameter, $\rho_a$ in the case of eq.\ (\ref{eq:Enun_app}), 
which can be interpreted as the 
centre-of-mass coordinate of the Pomeron state. In the following 
the dependence on this additional parameter will always be understood 
but often not be written out explicitly. 
In the high energy limit the leading contribution to the BFKL 
Pomeron comes from the state with vanishing conformal spin $n=0$, 
and only for this case the one-to-two Pomeron vertex has 
been calculated explicitly. 

One defines the one-to-two Pomeron vertex as 
\be
\label{eq:Vp2pdef}
V_{\pom\to2\pom}=
\delta_{a_1a_2}\,\delta_{a_3a_4}\,
\Big[(V_{2\to4}^{a_1a_2a_3a_4}D_2)(1,2,3,4)\Big]_{\Phi_2/\!\!/ E^{(\nu_c,n_c)} }
\otimes E^{(\nu_a,n_a)^*}(1,2)\,E^{(\nu_b,n_b)^*} (3,4)\,. 
\ee
The convolution $\otimes$ is defined as an integral 
$\int d\rho_1 \dots d\rho_4$ 
over the coordinates of all four outgoing gluons of the $2 \to 4$ vertex. 
Due to the complete symmetry of the vertex $V_{2\to4}$ in the 
four outgoing gluons it is not relevant which pairs of gluons are 
assumed to form Pomeron states. Here we have for simplicity 
chosen the pairs $(1,2)$ and $(3,4)$. 
The Pomerons are colour singlet states and the projection in colour 
space is performed via the contraction of the colour tensor of the 
vertex $V_{2\to4}$ with delta-tensors for the pairs of gluons 
corresponding to the Pomeron states. 
Note that we have chosen the outgoing Pomerons to be represented 
by complex conjugated eigenfunctions. 
We implicitly assume that $D_2$ is replaced 
by a pure BFKL amplitude according to (\ref{replaced2phi2}). 
As indicated in the subscript of the square bracket we further 
replace the full amplitude $\Phi_2$ by an eigenfunction 
$E^{(\nu,n)}$ of the BFKL kernel. The resulting quantity 
contains all relevant information of the coupling of the 
three Pomerons with the parameters $\{ \rho_i, h_i \} $, 
and the corresponding amplitude with three full Pomeron states 
is easily reconstructed. 
We will in the following always use the amplitude $D_2$ in the 
sense that it has to be replaced according to (\ref{replaced2phi2}), 
in particular also inside the function $G$. 

The projection (\ref{eq:Vp2pdef}) has been performed explicitly 
in \cite{Lotter:1996vk}. The resulting $1 \to 2$ Pomeron vertex 
was found to be a conformal three-point function of the three 
Pomeron coordinates, 
\be
V_{\pom\to2\pom}= C_{\{ h_i,\bar{h}_i \} }
\left[\rho_{ab}^{h_a+h_b-h_c} \rho_{bc}^{h_b+h_c-h_a} \rho_{ac}^{h_a+h_c-h_b} 
\rho_{ab}^{*\,\hbar_a+\hbar_b-\hbar_c} 
\rho_{bc}^{*\,\hbar_b+\hbar_c-\hbar_a} 
\rho_{ac}^{*\,\hbar_a+\hbar_c-\hbar_b} 
\right]^{-1}
\,,
\ee
where $\rho_a$ is the coordinate of the Pomeron state with 
quantum numbers $(\nu_a, n_a)$ etc., and 
the coefficient $C_{\{ h_i,\bar{h}_i \} }$ 
depends only on the conformal weights of the three Pomeron states. 
This finding motivates the symbolic notation of the vertex 
as a three-point correlation function, 
\be
V_{\pom\to2\pom}= 
\Big\langle E^{(\nu_a,n_a)^*}(\rho_a)\,E^{(\nu_b,n_b)^*}(\rho_b)\,
E^{(\nu_c,n_c)}(\rho_c) 
\Big\rangle
\,. 
\ee
The coefficient $C_{\{ h_i,\bar{h}_i \} }$ in the $1 \to 2$ Pomeron 
vertex is the sum of two terms, one of which is suppressed by two 
powers of $N_c$ with respect to the other. 
The numerical value of these two terms in the perturbative 
$1 \to 2$ Pomeron vertex was found in \cite{Korchemsky:1997fy} 
for the Pomeron `ground states' with $h=1/2$ which are leading 
in the high energy limit. 
The value of the term leading in $N_c$ obtained 
in this way starting from the EGLLA 
coincides exactly with the value found in 
\cite{Bialas:1997xp,Bialas:1997ig} in the dipole picture of high 
energy scattering which is based on the large-$N_c$ limit. 
In \cite{Bartels:2002au} the value of the perturbative 
$1 \to 2$ Pomeron vertex was later found to be 
of the same order of magnitude as  nonperturbative estimates 
of the triple-Pomeron vertex in Regge-type fits to scattering 
data at high energies. 

The formula for the $1 \to 2$ Pomeron vertex obtained in the 
dipole picture in \cite{Bialas:1997xp,Bialas:1997ig} exhibits a 
relatively simple structure. Motivated by this observation a 
generalisation of that formula to $1 \to n$ Pomeron vertices 
with arbitrary $n$ was conjectured in \cite{Peschanski:1997yx} 
motivated by results of \cite{Mueller:1994jq}. 
Interestingly, this generalised formula could be related to 
dual Shapiro-Virasoro amplitudes of a closed string theory. 
However, in \cite{Braun:1997nu} it was shown that in the 
dipole picture a direct transition, i.\,e.\ a transition local in 
rapidity, from one to more than two Pomerons is not possible. 
Instead, in the framework of the dipole picture such a transition 
can only occur via the iteration of the $1 \to 2$ Pomeron vertex, 
resulting in the well-known fan diagrams. The latter result 
does not necessarily exclude the existence of such vertices 
in general. Instead, it is well conceivable that it is the outcome 
of the large-$N_c$ limit used in the dipole picture. A natural 
next step beyond the limitations of the dipole model would be 
to include also higher multipoles. It is well conceivable 
that in such an extension of the dipole model (including 
quadrupoles for instance) a direct $1 \to 3$ Pomeron vertex 
would occur and possibly be of the form suggested in 
\cite{Peschanski:1997yx}. 

The perturbative $1\to3$ Pomeron vertex is defined in 
analogy to the $1 \to 2$ Pomeron vertex above, now using the 
$2 \to 6$ gluon vertex discussed in detail in the previous section. 
Due to the complete symmetry of that vertex in the six outgoing 
gluons it again does not matter which gluons are chosen to 
form Pomeron states. We choose the pairs $(1,2)$, $(3,4)$, 
and $(5,6)$, and the coordinates of these three Pomeron states 
will be $\rho_a$, $\rho_b$, and $\rho_c$, respectively. 
The Pomeron state attached to the vertex from above will 
have the coordinate $\rho_d$. 
We will often not write down the external coordinates as arguments, 
which coordinate is which will be clear from the conformal dimensions 
$\nu_a$, $\nu_b$, $\nu_c$, and~$\nu_d$. 

Hence the projection of the $2 \to 6$ gluon vertex defining the 
$1 \to 3$ Pomeron vertex can be written as 
\dmath2{
V_{\pom\to3\pom}&=&
\delta_{a_1a_2}\,\delta_{a_3a_4}\,\delta_{a_5a_6}\,
\Big[(V_{2\to6}^{a_1a_2a_3a_4a_5a_6}D_2)(1,2,3,4,5,6)\Big]_{\Phi_2/\!\!/E^{(\nu_d,n_d)}}
\otimes{}\cr&&{}\qquad\qquad\qquad
\otimes E^{(\nu_a,n_a)^*}(1,2)\,E^{(\nu_b,n_b)^*} (3,4)\,E^{(\nu_c,n_c)^*} (5,6)
\,.
&eq:Vp3pdef}%
Note again our convention in which the three outgoing Pomerons 
are described by complex conjugated eigenfunctions. 
Anticipating that the vertex will turn out to have the form of a conformal 
four-point function we will also use the symbolic notation of the vertex 
as a correlation function
\be
V_{\pom\to3\pom}= 
\Big\langle E^{(\nu_a,n_a)^*}(\rho_a)\,E^{(\nu_b,n_b)^*}(\rho_b)\,
E^{(\nu_c,n_c)^*}(\rho_c)\,E^{(\nu_d,n_d)}(\rho_d) 
\Big\rangle
\,. 
\ee
In the present paper we will restrict ourselves to the leading Pomeron 
states at high energies, i.\,e.\ the states with $n_i=0$, and this applies 
to all four Pomeron states in the $1 \to 3$ Pomeron vertex. 

The Pomeron vertices defined above are constructed from the number-changing 
vertices $V_{2 \to 4}$ and $V_{2 \to 6}$ of the EGLLA. It should be 
pointed out, however, that also other terms in the $n$-gluon amplitudes 
$D_n$ can in general give rise to multi-Pomeron vertices. 
In the case of the $1 \to 2$ Pomeron vertex for example also the 
reggeising part of the amplitude, i.\,e.\ the first term in 
(\ref{d4solutionpics}), gives a nonvanishing contribution 
upon projection on Pomeron states. So far this contribution 
has not yet been studied in much detail. 
In the case of the six-gluon amplitude $D_6$ there are two such 
reggeising terms, one being a superposition of two-gluon amplitudes 
and the other being a superposition of irreducible four-gluon amplitudes, 
containing in particular the effective transition vertices $I,J$ and $L$ of 
\cite{Bartels:1999aw}. 
We will not consider the contribution of these terms to the 
$1 \to 3$ Pomeron vertex in the present paper. These terms 
can well be important and clearly deserve further study. 
First steps in this direction have been performed in \cite{Volkerthesis}. 

\section{The $1\to3$ Pomeron vertex}
\label{sec:Wp3p}

\subsection{The colour structure}

The two gluons in a Pomeron form a colour singlet state, and hence 
the corresponding colour structure is $\delta_{ab}$.  Because the symmetric
structure constants $d_{abc}$ vanish when two of their indices are contracted 
less than half of the terms in the sum
in~(\ref{eq:V2to6}) contribute to the $1\to3$ Pomeron vertex.  The only
remaining terms are those in which the two colour indices of each Pomeron are
contracted with different $d$ tensors.  Since we chose as Pomerons 
the pairs (12), (34) and (56) of gluons, 
this leaves the following four permutations of
(colour, momentum or coordinate) indices:
\dmath2{
(1,3,5;2,4,6),\quad(1,3,6;2,4,5),\quad(1,4,5;2,3,6),\hbox{ and }(1,4,6;2,3,5)
\,.&&&eq:projind}

Each of these permutations gives rise to the same colour factor because of the
symmetry of $d_{abc}$.  Since each $d_{abc}$ tensor is contracted with one
index of each of the $\delta$s, the result is the contraction of two $d$
tensors. For general $N_c$ we denote this contraction by $\cconst$. 
Its value is, calculated for example for the first permutation, 
\be
\cconst = 
d_{a_1a_3a_5}d_{a_2a_4a_6}\delta_{a_1a_2}\delta_{a_3a_4}\delta_{a_5a_6}=
d_{a_1a_3a_5}d_{a_1a_3a_5}=\frac1{N_c}(N_c^2-4)(N_c^2-1)
\,.
\ee

\subsection{The spatial part}
\label{sec:p3pspatial}

\subsubsection{Introduction}

The lion's share of the work to be done for the projection is sorting out all
the terms of the spatial part of $V_{2\to6}$.  This work is complicated by the
fact that the simple representation of $V_{2\to6}$ in terms of $G$ does not 
help in obtaining the $1\to3$ Pomeron vertex. 
That is because due to $\Ez{}(\rho,\rho)=0$ some of the component 
functions of $G$ vanish on projection, others don't. 

The brute-force way of extracting those that remain would be to write out
$(WD_2)$ in terms of the functions $a$, $b$, $c$, $s$ and~$t$ and sort out those
that vanish upon projection. However, there is a more
systematic way.  First, we observe that the four permutations
(\ref{eq:projind}) remaining after the contraction of colour tensors all lead
to the same term.  This is because they differ only by swapping two coordinates
of the same Pomeron wave function, which is symmetric.  Hence it is sufficient
to consider only one of the permutations and multiply the result by four.

The second important observation is that the conformal eigenfunctions~$\Ez{}$
vanish if their two coordinate arguments are equal, $\Ez{}(\rho,\rho)=0$. 
Therefore terms in which
both argument indices of the same Pomeron occur in the same argument of~$G$
vanish under projection.  This is the key to finding out which terms remain
after projection.  Let us now have a look at the first permutation
in~(\ref{eq:projind}).  The spatial part of the vertex for this permutation has
the following form:
\dmath2{
(WD_2)(1,3,5;2,4,6)&=&
\frac{g^4}8 \sum_{M\in\mathcal P(\{1,\ldots,6\})}\mskip -20mu
(-1)^{\#M}\;G(135\backslash M,M,246\backslash M)\,.
&eq:WD2_135}%
We can see immediately that a term from this sum vanishes if two coordinates of
the same Pomeron are contained in the middle argument of~$G$.  The first and
last argument of~$G$ contain only coordinates or momenta belonging to different
Pomerons and hence cannot cause the term to vanish under projection.  Therefore
all the terms for which $M$ does not contain both indices of a Pomeron remain:
\dmath2{
(WD_2)_{\pom\to3\pom}(1,3,5;2,4,6)&=&
\frac{g^4}8 
\mskip -30mu
\sum_{{M\in\mathcal P(\{1,\ldots,6\}) \atop 
M\not\supset\{1,2\}\wedge M\not\supset\{3,4\}\wedge M\not\supset\{5,6\}}} 
\mskip -60mu
(-1)^{\#M}\;G(135\backslash M,M,246\backslash M)\,.
&eq:WD2proj}%
In words: 
$M$ may contain either 1 or 2 or none of these two indices,
and 3 or 4 or neither of them, and 5 or 6 or neither.  This yields
$3\cdot3\cdot3=27$ terms.  However, two of these vanish because of a property
of the $G$ function: It vanishes when its first or third argument is empty, see
Eq.~(\ref{eq:G-23}).  This eliminates the terms for $M=\{1,3,5\}$ and
$M=\{2,4,6\}$.  Hence 25 terms remain.

The function $G$ is composed of functions with one, two and with three
arguments.  The function with one argument, $c$, does not contribute at all to
the Pomeron vertex.  Since its argument always contains all indices, $c$ does
not survive the projection.  The functions with two arguments, $b$ and $t$, are
also components of the BFKL kernel.  We will discuss them in the next part of
this section.  The functions with three arguments, $a$ and $s$, will be dealt
with after that. 

\subsubsection{The BFKL term}

The functions $b$ and $t$ survive only if they are convoluted with Pomerons in
one specific way, namely if each Pomeron's arguments are convoluted with
different arguments of $b$ or~$t$.  This means that both arguments of $b$
resp.\ $t$ contain three indices, one from each Pomeron.  First of all, this is
the case for the term in (\ref{eq:WD2}) where $M=\varnothing$.  

Besides, there are terms in the sum~(\ref{eq:WD2proj}) where $M$ contains only
indices from the first or only from the second group of arguments of~$(WD_2)$.
Since one argument of $b$ and $t$ contains two arguments of $G$ taken together
(see~(\ref{eq:G_symb})), one of the two $b$ and $t$ functions would have
momenta of different Pomerons in each argument and hence would contribute.
However, by replacing $M$ by its complement with respect to the group of
arguments of $(WD_2)$, an identical term with opposite sign can be obtained
(unless $M$ is identical to the whole group of arguments, but these terms
vanish, see above).  Therefore all such terms cancel out.

To illustrate this, let us look at an example.  We again consider the first of
the permutations in~(\ref{eq:projind}) with the spatial
function~(\ref{eq:WD2_135}).  For example, $M=\{1,3\}$ contains only indices
from the first three arguments of~$(WD_2)$.  The corresponding term in the sum
contains the following $G$ function:
\dmath{0.5}{
G(5,13,246)&=&
\frac{g^2}2\big[2\,c(123456)-2\,b(135,246)-2\,b(12346,5)+2\,a(13,5,246)
\cr&&\quad
{}+t(135,246)+t(12346,5)-s(13,5,246)-s(13,246,5)\big]\,.
&eq:G_5_13}%
The terms with more than three indices in one argument, such as $c$ and the
second $b$ and~$t$ functions, vanish because this identifies two coordinates of
a Pomeron.  For now, we ignore the functions $a$ and~$s$; we will deal with
them in the next section.  The remaining terms, $-2\,b(135,246)$ and
$t(135,246)$ do not vanish.  However, the same terms occur in the term for the
complement of $M$, $M'=\{1,3,5\}\backslash M=\{5\}$.  The corresponding $G$
function is:
\dmath{0.5}{
G(13,5,246)&=&
\frac{g^2}2\big[2\,c(123456)-2\,b(135,246)-2\,b(2456,13)+2\,a(5,13,246)
\cr&&\quad
{}+t(135,246)+t(2456,13)-s(5,13,246)-s(5,246,13)\big]\,.
&eq:G_13_5}%
Again, $-2\,b(135,246)$ and $t(135,246)$ remain.  But this term has the
opposite sign because of the prefactor~$(-1)^{\#M}$.  So the $t$ and $b$
functions for $M\not=\varnothing$ cancel out.\footnote{The deeper reason for
this is that $6/2=3$ is odd.  It can be conjectured that the same mechanism of
cancellation occurs in the projection of higher $1\to n$ Pomeron vertices from
$2\to2n$ reggeised gluon vertices for odd~$n$.}

\begin{figure}[ht] 
\centre{\input{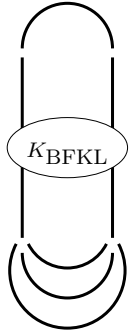}}
\caption{Graphical representation of the part of the $1\to3$ Pomeron vertex 
                which contains the BFKL kernel. The half-circular lines indicate 
                Pomeron wave functions.}
\label{fig:p3pbfkl}
\end{figure}

Only for $M=\varnothing$ the functions $b$ and~$t$ contribute to the $1\to3$
Pomeron vertex.  Knowing that $G$ becomes a derivative of the BFKL kernel when
its middle momentum argument vanishes (see Eq.\ (\ref{eq:G_BFKL})), we can
immediately write out this term.  Since there is one such term for each of the
four permutations whose colour structure does not vanish, we have to multiply
the result by~4.  We now replace the full amplitude $\Phi_2$ in $G$ 
by the eigenfunction $E^{(\nu_d,0)}$ of the BFKL kernel as discussed in section 
\ref{sec:pvertices} and obtain the following BFKL term:
\dmath2{
V_{\pom\to3\pom}^{BFKL}&=&
\cconst \frac{g^4}8 4
\int d^2\rho_1\,d^2\rho_2\,
\Ez a^*(\rho_1,\rho_2)\,
\Ez b^*(\rho_1,\rho_2)\,
\Ez c^*(\rho_1,\rho_2)\cdot{}
\crd&&\qquad\qquad\qquad
{}\cdot\frac1{N_c}
\Delta_{1}\,\Delta_{2}\,\big(K_\bfklind\otimes \Ez d\big)(\rho_1,\rho_2)
\cr&=&
\cconst \frac{g^4}8 4
\frac{\chi(\nu_d,0)}{N_c}
\,(4\,\nu_d^2+1)^2
\int \frac{d^2\rho_1\,d^2\rho_2}{|\rho_{12}|^4}
\Ez a^*(\rho_1,\rho_2)\,
\Ez b^*(\rho_1,\rho_2)
\cdot{}\crd
&&\qquad\qquad\qquad\qquad\qquad\quad{}\cdot
\Ez c^*(\rho_1,\rho_2)\,
\Ez d(\rho_1,\rho_2)
\cr&=&
\cconst \frac{g^6}{16} 4 \frac2{(2\pi)^3}
\xi(\nu_d)\,(4\,\nu_d^2+1)^2
\;\int \frac{d^2\rho_1\,d^2\rho_2}{|\rho_{12}|^4}
\cEz12a\cdot{}\crd
&&\qquad{}\cdot
\cEz12b\cEz12c\eEz12d\,.
&eq:p3pbfkl}%
We have evaluated the convolution with the BFKL kernel using the BFKL
eigenvalue
\dmath2{
\int d^2\rho_1'\,d^2\rho_2'\; K_\bfklind(\rho_1,\rho_2;\rho_1',\rho_2')
\;E^{(\nu,n)}(\rho_1',\rho_2')&=&
\chi(\nu,n)\;E^{(\nu,n)}(\rho_1,\rho_2)\,,
&eq:BFKLeigeneq}%
and put in the derivative of the wave function~(\ref{eq:LapLapE}). 
In addition, in the last step 
we have rewritten the BFKL eigenvalue $\chi$ as a function
$\xi$ which was already used in~\cite{Lotter:1996vk}.  It differs from $\chi$ only
in that it does not contain factors of $N_c$ and the coupling constant:
\dmath2{
\xi(\nu)&=&2\pi
\left[2\,\psi(1)-\psi\!\left(\frac12+i\nu\right)
-\psi\!\left(\frac12-i\nu\right)\right]
=\frac{8\pi^3}{N_c g^2}\chi(\nu,0)\,.
&eq:xi}%
Figure~\ref{fig:p3pbfkl} shows a graphical representation of this part of the
vertex.

\subsubsection{The $\alpha$ terms}

\begin{figure}[ht] 
\centre{\Large\input{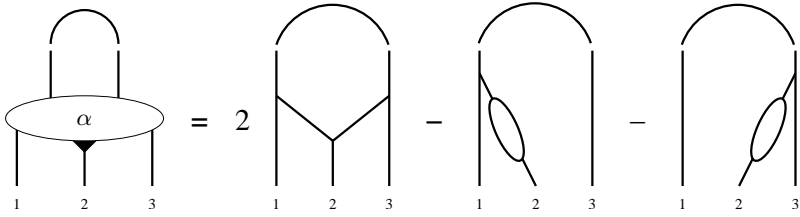}}
\caption{Graphical representation of Eq.~(\ref{eq:alpha}).  The arcs
                represent Pomeron wave functions defined into the functions $a$, $s$
                and $\alpha$.}
\label{fig:alpha}
\end{figure}

What remains now are the functions with three arguments, $a$ and~$s$.  Since
they have the same properties (both vanish only when their first argument, 
i.\,e.\ the second argument of $G$, is empty), they always occur together, and the 
result of the projection can be expressed in terms of the function $\alpha$ 
defined as 
\dmath0{
\alpha(2,1,3)&=&2a(2,1,3)-s(2,1,3)-s(2,3,1)\,.
&eq:alpha}%
$\alpha$ is symmetric in its last two arguments.  Figure~\ref{fig:alpha} shows
a graphical representation of this equation using the diagrammatic notation 
used for example in \cite{Bartels:1999aw}. The middle leg of $\alpha$, which
corresponds to its first argument, is marked to remind us of the fact that $\alpha$ 
is symmetric in the other two arguments, but not this one.

We will now divide the remaining terms from (\ref{eq:WD2proj}) into groups that
lead to the same type of integral after convolution with the wave functions.
There are 24 terms since the case $M=\varnothing$ has already been dealt with
above.  The most obvious classification criterion is the number of elements $M$
contains, i.\,e.\ the number of momenta or coordinates in the first argument of
$\alpha$.  There are six terms for which $M$ has one element, twelve for which
it has two and six with three elements.  (Normally there would be eight in the
last class, but for two of them $M$ contains a whole group of arguments and $G$
vanishes, see above.)  Not that it is quite natural that the numbers of terms in all classes
are divisible by six.  Six is the number of permutations of three objects.
Because of the high symmetry of the $2\to6$ reggeised gluon vertex, there are
terms with every Pomeron coupling to every pair of arguments of the
function~$G$.  Terms which differ only by permutations of the labels
($={}$external coordinates) of the Pomerons belong to the same class since they
lead to the same type of integral.  The terms from different classes contribute
with different signs due to the factor $(-1)^{\#M}$ in~(\ref{eq:WD2_135}).  The
second class has positive sign, the others negative sign.

\begin{figure}[ht] 
\centre{\Large\input{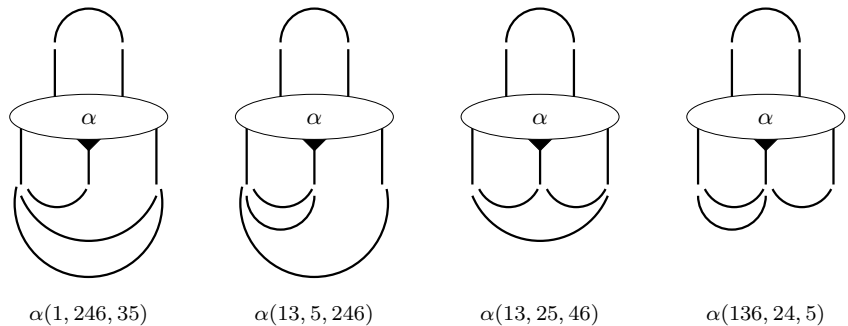}}
\caption{Graphical representation of the classes of terms with the $\alpha$ 
                function.  One representative term is printed underneath each graph
                (assuming the Pomerons (12), (34) and (56)).  The half-circular lines 
                indicate Pomeron wave functions.}
\label{fig:p3palpha}
\end{figure}

The second of the classes presented above can be subdivided further.  It
contains those terms for which $M$, and hence the first argument of $\alpha$,
contains two indices.  It makes a difference whether the second and third
arguments also contain two or whether one has three and the other only one.
With this subdivision, we have four classes in total, with six terms each.
There is an intuitive interpretation of these classes which is presented
graphically in Figure~\ref{fig:p3palpha}.  The classes differ in how many
Pomerons are attached to the different legs of the $\alpha$ function. 

Bearing in mind that $\alpha$ is symmetric in its last two arguments but not
its first, it is easy to see that the four classes are indeed disjoint.  What
is more, they are also complete.  It is impossible to construct a term which
does not belong to one of the classes (taking into account the properties of 
$\alpha$ and the Pomeron amplitude). 

\subsubsection{A closer look at the $\alpha$ terms}

Before writing down the integrals containing $\alpha$ which make up the largest
part of the $1\to3$ Pomeron vertex, let us write down the configuration space
form of $\alpha$:
{\def\mbf{} 
\dmath2{
\alpha(\rho_2,\rho_1,\rho_3)
&=&\frac2{(2\pi)^3}\bigg[
{-}\Delta_2\,K_0(m|\mbf\rho_{12}|)\,K_0(m|\mbf\rho_{23}|)\,
-\left(2\pi\,\delta^2(\mbf\rho_{12})\,K_0(m|\mbf\rho_{12}|)
        -\frac1{|\mbf\rho_{12}|^2}\right)
\crd&&\qquad\qquad
-\left(2\pi\,\delta^2(\mbf\rho_{23})\,K_0(m|\mbf\rho_{23}|)
        -\frac1{|\mbf\rho_{23}|^2}\right)
\bigg]
\Delta_1\,\Delta_3\,\Phi_2(\mbf\rho_1,\mbf\rho_3)
\cr
&=&\frac2{(2\pi)^3}\bigg[
-\bigg(-2\pi\,\delta^2(\mbf\rho_{12})\,K_0(m|\mbf\rho_{23}|)
+\frac1{\rho_{12}\rho_{23}^*}+\frac1{\rho_{12}^*\rho_{23}}
\crd&&\qquad\qquad\quad\ {}
-2\pi\,\delta^2(\mbf\rho_{23})\,K_0(m|\mbf\rho_{12}|)\bigg)
-2\pi\,\delta^2(\mbf\rho_{12})\,K_0(m|\mbf\rho_{12}|)
\crd&&\qquad\quad\ {}
-2\pi\,\delta^2(\mbf\rho_{23})\,K_0(m|\mbf\rho_{23}|)
+\frac1{|\mbf\rho_{12}|^2}+\frac1{|\mbf\rho_{23}|^2}
\bigg]
\Delta_1\,\Delta_3\,\Phi_2(\mbf\rho_1,\mbf\rho_3)
\cr
&=&\frac2{(2\pi)^3}\bigg[
-2\pi\big(\delta^2(\mbf\rho_{12})-\delta^2(\mbf\rho_{23})\big)
\big(K_0(m|\mbf\rho_{12}|)-K_0(m|\mbf\rho_{23}|)\big)
\crd&&\qquad\qquad
+\left|\frac1{\rho_{12}}+\frac1{\rho_{23}}\right|^2
\bigg]
\Delta_1\,\Delta_3\,\Phi_2(\mbf\rho_1,\mbf\rho_3)
\cr
&=&\frac2{(2\pi)^3}\bigg[
2\pi\big(\delta^2(\rho_{12}){-}\delta^2(\rho_{23})\big)
\ln\frac{|\rho_{12}|}{|\rho_{23}|}
{+}\left|\frac{\rho_{13}}{\rho_{12}\rho_{23}}\right|^2
\bigg]
\Delta_1\Delta_3\Phi_2(\rho_1,\rho_3)\,, 
&eq:alpha_c}}%
where we have made use of the identity 
$\del (1/\rho^*) = \del^* (1/\rho) = \pi \delta^2(\rho)$. 

The two delta functions identify the second or third coordinate argument of $G$
with its first.  When this function is convoluted with further Pomeron wave
functions as illustrated in Figure~\ref{fig:p3palpha}, one or both of the delta
function terms vanish after integration because the delta functions identify
two coordinates of a Pomeron.  For the last two groups in that Figure 
both delta function
terms vanish (since there are Pomerons between the middle and both outer legs
of $\alpha$), for the others only one.

Several terms of the function $\alpha$ are potentially divergent under an
integral.  The terms $\delta^2(\rho_{12})\,\ln|\rho_{12}|$ are obviously
dangerous, but also the fraction may cause a logarithmic divergence with
respect to the integration over $\rho_{12}$ or~$\rho_{23}$.  The Pomeron wave
function for zero conformal spin $\Ez{}(\rho_1,\rho_2)$ contains one power of
$|\rho_{12}|$ and can hence regularise these divergences.  Eliminating the
delta functions requires that a Pomeron is attached between the middle and one
of the outer legs of~$\alpha$.  Therefore the terms from the last two groups in
Figure~\ref{fig:p3palpha} are finite while in the others one divergence remains
and has to be regularised.

We will now write down the integrals resulting from each of the four groups of
terms.  We will denote them (and the groups) with an upper index equal
to~$\#M$, the number of elements in~$M$.  The two groups with $\#M=2$ will be
distinguished by the minimal number of coordinates in the other arguments of
the $G$ function and denoted by ``(2,1)'' and ``(2,2)''.

The trajectory function contains two terms which are ultraviolet divergent
under an integral and have to be regularised: $1/|\rho|^2$ and
$\delta(\rho)\ln\rho$.  We regularise the fraction $1/|\rho|^2$ with a theta
function $\theta(|\rho|-\epsilon)$, i.\,e.\ a UV cutoff.  The term
$\delta^2(\rho)\ln|\rho|$ is regularised by replacing $\ln|\rho|$ with
$\ln\epsilon$.

For the first group of terms ($\#M=1$), we obtain 
\dmath2{
V_{\pom\to3\pom}^{(1)}&=&
-\cconst \frac{g^6}{16} 4 \sumabc
\int d^2\rho_1\,d^2\rho_2\,d^2\rho_3\,d^2\rho_4\,d^2\rho_5\,d^2\rho_6\,
\Ez a^*(\rho_1,\rho_2)\,\Ez b^*(\rho_3,\rho_4)\cdot{}
\crd&&\qquad\qquad
{}\cdot
\Ez c^*(\rho_5,\rho_6)
\;\delta^2(\rho_{24})\,\delta^2(\rho_{26})\,\delta^2(\rho_{35})\;
\frac2{(2\pi)^3}
\bigg[2\pi\,\delta^2(\rho_{13})\,\ln\frac\epsilon{|\rho_{12}|}
\crd&&\qquad\qquad{}
+\frac{|\rho_{23}|^2}{|\rho_{12}|^2|\rho_{13}|^2}\theta(|\rho_{13}|-\epsilon)
\bigg]\,\Delta_2\,\Delta_3\,\Ez d(\rho_2,\rho_3)
\cr&=&
-\cconst \frac{g^6}{16} 4
 \frac2{(2\pi)^3}\sumabc\int d^2\rho_1\,d^2\rho_2\,d^2\rho_3\,
\Ez a^*(\rho_1,\rho_2)\,\Ez b^*(\rho_2,\rho_3)\cdot{}
\crd&&\qquad\qquad
{}\cdot
\Ez c^*(\rho_2,\rho_3)\,\left[2\pi\,\delta^2(\rho_{13})\,\ln\frac\epsilon{|\rho_{12}|}
+\frac{|\rho_{23}|^2}{|\rho_{12}|^2|\rho_{13}|^2}\theta(|\rho_{13}|-\epsilon)
\right]\cdot{}
\crd&&\qquad\qquad
{}\cdot
\Delta_2\,\Delta_3\,\Ez d(\rho_2,\rho_3)\,.
&eq:Vp3p1_step1}%
The constant $\cconst$ is the colour factor, the factor 4 the combinatorial
factor from the four permutations~(\ref{eq:projind}).  The factor $g^6/16$ is
the product of the prefactors in $(WD_2)$ and~$G$.  We have rewritten the sum
over the six terms of this group as a sum over the permutations of the Pomeron
indices which is equivalent to it by way of renaming the integration variables.
It is denoted by $\sumabc$.  We can now insert the explicit form of the Pomeron
wave functions.  This allows us to evaluate the double Laplacian 
calculated in the Appendix (see Eq.~(\ref{eq:LapLapE})), and we find 
\dmath3{
V_{\pom\to3\pom}^{(1)}&=&\ldots =
-\cconst \frac{g^6}{16} 4 \frac2{(2\pi)^3}(4\,\nu_d^2+1)^2
\sumabc\int d^2\rho_1\,d^2\rho_2\,d^2\rho_3\,\cdot
\crd&&\qquad{}\cdot
\cEz12a\,\cEz23b\,\cEz23c\cdot{}
\crd&&\qquad{}\cdot
\left[2\pi\,\delta^2(\rho_{13})\,\ln\frac\epsilon{|\rho_{12}|}
+\frac{|\rho_{23}|^2}{|\rho_{12}|^2|\rho_{13}|^2}\theta(|\rho_{13}|-\epsilon)
\right]
\frac1{|\rho_{23}|^4}\eEz23d
\cr&=&
-\cconst \frac{g^6}{16} 4 \frac2{(2\pi)^3}(4\,\nu_d^2+1)^2
\sumabc\int \frac{d^2\rho_2\,d^2\rho_3}{|\rho_{23}|^4}
\cEz23b\cdot{}
\crd&&{}\cdot
\cEz23c\,\eEz23d\int d^2\rho_1\,
\bigg[2\pi\,\delta^2(\rho_{13})\,\ln\epsilon
\crd&&{}
+\frac{|\rho_{23}|^2}{|\rho_{12}|^2|\rho_{13}|^2}
\theta\!\left(\frac{|\rho_{13}|}{|\rho_{12}|}-\epsilon\right)\bigg]
\,\cEz12a
&eq:p3p1}%
We have converted a logarithm into a factor inside the theta function.  This
has the advantage that now the theta function is invariant under dilatations.
It is done according to the following identity:
\dmath2{
&&\lim_{\epsilon\to0}\int d^2\rho_1
\left(
\frac{\theta(|\rho_{13}|-\epsilon/\lambda)}{|\rho_{13}|^2}f(\rho_1)
-\frac{\theta(|\rho_{13}|-\epsilon)}{|\rho_{13}|^2}f(\rho_1)
\right)=
\cr&&\qquad\qquad=
\lim_{\epsilon\to0}\int d^2\rho_1
\frac{\theta(|\rho_{13}|-\epsilon/\lambda)-\theta(|\rho_{13}|-\epsilon)}
{|\rho_{13}|^2}
\big(f(\rho_3)+\mathcal O(\epsilon)\big)
\cr&&\qquad\qquad=
2\pi\,f(\rho_3)\,\lim_{\epsilon\to0}\int\limits_{\epsilon/\lambda}^{\epsilon}
d|\rho_{13}|\frac1{|\rho_{13}|} 
\cr&&\qquad\qquad=
2\pi\,\ln\lambda\,f(\rho_3)
\cr&&\qquad\qquad=
2\pi\,\ln\lambda\int d^2\rho_1\,\delta^2(\rho_{13})\,f(\rho_1)\,.\qquad
&eq:tregchange}%
Here we have used it for $\lambda=1/|\rho_{12}|$ and $f(\rho_1)=\Ez
a^*(\rho_1,\rho_2)$.  One can argue~\cite{Volkerthesis} 
that (\ref{eq:tregchange}) can
be applied even when the factor $\lambda$ contains the integration variable.

It is easily shown that $V_{\pom\to3\pom}^{(1)}$ alone has the
transformation properties of a conformal four-point function.  Invariance under
translations and rotations is trivial.  Invariance of the integral operator
under dilatation with a factor $\lambda$ is not hard to prove: The three
integrals give a factor $\lambda^6$, the denominator $|\rho_{23}|^4$ a
factor~$\lambda^{-4}$.  Both the delta function and the fraction in the
rectangular brackets give a factor~$\lambda^{-2}$.

There remains inversion.  The first two integrals together with the denominator
are invariant.  So is the $\rho_1$ integral together with the delta function
and the fraction.  However, the theta function gives rise to an extra
logarithmic term inside the brackets.  Using relation~(\ref{eq:tregchange})
with respect to the $\rho_1$ integration for $\lambda=|\rho_2/\rho_3|$, it is
computed to $|\rho_{23}|^2/|\rho_{12}|^2 \cdot
2\pi\,\delta(\rho_{13})\ln|\rho_2/\rho_3|$.  After performing the $\rho_1$
integration, everything except this logarithm is symmetric under the exchange
$\rho_2\leftrightarrow\rho_3$.  Hence the additional term is overall
antisymmetric and vanishes under the integration over $\rho_2$ and $\rho_3$.

The integrals resulting from the other groups of terms are derived analogously.
In all cases, the remaining integration variables are $\rho_1$, $\rho_2$ and
$\rho_4$, which we will rename~$\rho_3$.  The second group, for which $\#M=2$
and one of the arguments of $G$ contains only one coordinate, also requires
regularisation.  We deal with that as shown for $V_{\pom\to3\pom}^{(1)}$,
including absorbing the logarithm into the theta function.  Like
$V_{\pom\to3\pom}^{(1)}$, the vertices belonging to the other groups are also
conformal four-point functions on their own.  Here they are:
\dmath3{
V_{\pom\to3\pom}^{(2,1)}&=&
\cconst \frac{g^6}{16} 4 \frac2{(2\pi)^3}(4\,\nu_d^2+1)^2
\sumabc \int d^2\rho_1\,d^2\rho_2\,d^2\rho_3\,
\frac1{|\rho_{12}|^4}
\cEz12a\cdot{}
\crd&&
{}\cdot\cEz13b\cEz13c\bigg[2\pi\,\delta^2(\rho_{23})\ln\epsilon
\crd&&{}
+\frac{|\rho_{12}|^2}{|\rho_{13}|^2|\rho_{23}|^2}
\theta\!\left(\frac{|\rho_{23}|}{|\rho_{13}|}-\epsilon\right)\bigg]
\eEz12d
&eq:p3p21\cr
V_{\pom\to3\pom}^{(2,2)}&=&
\cconst \frac{g^6}{16} 4 \frac2{(2\pi)^3} (4\,\nu_d^2+1)^2
\sumabc\int d^2\rho_1\,d^2\rho_2\,d^2\rho_3\,
\frac1{|\rho_{12}|^2|\rho_{13}|^2|\rho_{23}|^2}
\cdot{}
&eq:p3p22\cr\crd
&& {}\cdot\cEz12a\cEz13b\cEz23c\eEz23d
\cr
V_{\pom\to3\pom}^{(3)}&=&
-\cconst \frac{g^6}{16} 4 \frac2{(2\pi)^3} (4\,\nu_d^2+1)^2
\sumabc\int d^2\rho_1\,d^2\rho_2\,d^2\rho_3\,
\frac1{|\rho_{12}|^2|\rho_{13}|^2|\rho_{23}|^2}
\cdot{}
&eq:p3p3\cr\crd
&& {}\cdot\cEz12a\cEz13b\cEz13c\eEz23d\,. 
}

\subsection{Simplifying the spatial part and the function $\Psi$}

\subsubsection{The $1\to3$ Pomeron vertex as a conformal four-point function}

In the previous section we have learnt that each of the five terms of the
$1\to3$ Pomeron vertex is a conformal four-point function.  This means that
they are severely constrained in their form.  The general form can be
simplified even further because we use only ground state wave functions for 
the Pomerons, so that the
two conformal weights are equal: $h=\bar h=\frac{1}2{} + i\nu$. 
Besides, we have to take
into account that we have used complex conjugated wave functions for the
Pomerons attached to the vertex from below.  This is equivalent to replacing
$\nu\to-\nu$.  The resulting formula for our four-point function reads:
\dmath2{
\bigg\langle \Ez a^*(\rho_a)\,\Ez b^*(\rho_b)\,&&\Ez c^*(\rho_c)\,\Ez d(\rho_d) 
\bigg\rangle=
\crd
&&=\Psi(x,x^*;\{\nu_i\})\prod_{i<j}
|\rho_{ij}|^{-\frac{2}{3}+
2i\left(-\tilde\nu_i-\tilde\nu_j+\frac13\sum_k \tilde\nu_k\right)}
\,,\qquad\quad\quad
&eq:our4point\cr
&&\qquad\qquad x=\crossrat abcdacbd\,,&eq:crossratx}%
where $\tilde\nu_i$ is $-\nu_i$ for $i=a,b,c$ and $\nu_d$ for $i=d$.  The
sum over $k$ in the exponent runs over the four indices $a, \dots, d$. 
The single argument of the conformal eigenfunctions is the external coordinate of
the Pomeron state.  We have omitted the coordinates of the reggeised gluons
contained in the Pomerons since they are integrated over to obtain the
correlation function on the left-hand side.  By writing down all the factors, one
can obtain the following explicit form of the four-point function:
{
\def\fourrat#1#2#3#4#5#6#7#8#9{
\left(\frac{|\rho_{a#1}|^2|\rho_{#2#3}|^2|\rho_{#4d}|^2}
{|\rho_{#5#6}||\rho_{#5#7}||\rho_{#6#7}|}\right)^{-\frac13#8\frac23i\nu_{#9}}}
\dmath2{
\Big\langle \Ez a^*(\rho_a)&&\,\Ez b^*(\rho_b)\,\Ez c^*(\rho_c)\,\Ez d(\rho_d) 
\Big\rangle ={}&eq:our4exp\cr
{}=\Psi&&(x,x^*;\{\nu_i\})
\fourrat bacabcd+a  \fourrat bbcbacd+b  \cdot{}
\cr&&\qquad\qquad{}\cdot
\fourrat cbccabd+c \fourrat dbdcabc-d\,.
}
}%
Now we rewrite the ratios in brackets as follows (shown here for the first
expression):
\dmath2{
\frac{|\rho_{ab}|^2|\rho_{ac}|^2|\rho_{ad}|^2}
{|\rho_{bc}||\rho_{bd}||\rho_{cd}|}&=&
\left|\crossrat abcdacbd\right|
\left|\crossrat abcdadbc\right|
\frac{|\rho_{ac}|^3|\rho_{ad}|^3}{|\rho_{cd}|^3}\,.
&eq:rewrite4frac}%
The first of the anharmonic ratios appearing here is what we defined as $x$ in
Eq.~(\ref{eq:our4point}), the second is~$x/(1-x)$.  We rewrite the following
two factors in (\ref{eq:our4exp}) analogously to (\ref{eq:rewrite4frac}), 
i.\,e.\ with the indices $a$, $b$ and $c$ permutated in order: $a\to b\to c$.  
Taking these three expressions together, one notices that the constant part of the
exponents, $-\frac13$, of the first three anharmonic ratios cancel.  From the
last factor in~(\ref{eq:our4exp}), we pull out the two different anharmonic
ratios which have $\rho_{ad}\rho_{bc}$ in the numerator, again analogously
to~(\ref{eq:rewrite4frac}). This leads to 
{\def\fourthree#1#2#3#4#5#6#7#8{%
{\left(\frac{|\rho_{#1#2}||\rho_{#3#4}|}{|\rho_{#5#6}|}\right)^{-1#72i\nu_{#8}}
}}
\dmath2{
\Big\langle\ldots\Big\rangle&=&
\Psi(x,x^*;\{\nu_i\})\;
\left|\crossrat abcdacbd\right|^{\frac23i(\nu_a-\nu_c)}
\left|\crossrat abcdadbc\right|^{\frac13+\frac23i(\nu_a-\nu_b+\nu_d)}
\left|\crossrat adbcacbd\right|^{-\frac13+\frac23i(\nu_b-\nu_c-\nu_d)}
\cdot\crd&&\hskip-3mm{}\cdot
\fourthree acadcd+a \fourthree abbdad+b \fourthree bccdbd+c \fourthree bdcdbc-d
\cr&=&
\Psi(x,x^*;\{\nu_i\})\;\;
|x|^{\frac13+\frac23i(2\nu_a-\nu_b-\nu_c+\nu_d)}\;
|1-x|^{-\frac23+\frac23i(-\nu_a+2\nu_b-\nu_c-2\nu_d)}
\cdot{}&eq:fourpointderive\cr\crd&&\hskip-3mm{}\cdot
\fourthree acadcd+a \fourthree abbdad+b \fourthree bccdbd+c \fourthree bdcdbc-d
}}%
Given a conformal four-point function, we can obtain the corresponding $\Psi$
function by solving~(\ref{eq:our4point}) for~$\Psi$.  After expressing the
powers of~$|\rho_{ij}|$ in terms of $x$ and of ratios of three $|\rho_{ij}|$s,
we obtain the formula
\dmath2{
\Psi&=&
|x|^{-\frac13+\frac23i(-2\nu_a+\nu_b+\nu_c-\nu_d)}\;
|1-x|^{\frac23+\frac23i(\nu_a-2\nu_b+\nu_c+2\nu_d)}
\cdot{}\cr&&{}\cdot
\psithree acadcd-a \psithree abbdad-b \psithree bccdbd-c \psithree bdcdbc+d
\cdot{}
\cr
&&{}\cdot
\Big\langle \Ez a^*(\rho_a)\,\Ez b^*(\rho_b)\,\Ez c^*(\rho_c)\,\Ez d(\rho_d) 
\Big\rangle\,.
&eq:getpsi}%
From now on we will not write the arguments of $\Psi$ explicitly any more.  It
depends on the $\nu_i$ as well as on $x$ and~$x^*$.

\subsubsection{The BFKL term}

Since each of the integral expressions derived in Section~\ref{sec:p3pspatial}
is a conformal four-point function, the $\Psi$ function of the vertex can be
written as a sum over five $\Psi$ functions associated with the BFKL term and
the four $\alpha$ terms.  They can be computed by inserting the integrals into
Eq.~(\ref{eq:getpsi}).  We will start by calculating the $\Psi$ function of
the BFKL term,~$\Psi_\bfklind$.  Inserting (\ref{eq:p3pbfkl}) into the formula
for $\Psi$, we get:
\dmath2{
\Psi_\bfklind&=&
\cconst \frac{g^6}{16} 4 \frac2{(2\pi)^3} \xi(\nu_d)\,(4\nu_d^2+1)^2\;
|x|^{-\frac13+\frac23i(-2\nu_a+\nu_b+\nu_c-\nu_d)}\;
\cdot{}\crd&&{}\cdot
|1-x|^{\frac23+\frac23i(\nu_a-2\nu_b+\nu_c+2\nu_d)}
\psithree acadcd-a \psithree abbdad-b 
\cdot{}\crd&&{}\cdot
\psithree bccdbd-c \psithree bdcdbc+d
\int \frac{d^2\rho_1\,d^2\rho_2}{|\rho_{12}|^4}
\cEz 12a
\cdot{}\crd&&{}\cdot
\cEz 12b\,\cEz 12c\,\eEz 12d
\cr&=&
\cconst \frac{g^6}{16} 4 \frac2{(2\pi)^3} \xi(\nu_d)\,(4\nu_d^2+1)^2\;
|x|^{-\frac13+\frac23i(-2\nu_a+\nu_b+\nu_c-\nu_d)}\;
\cdot{}\crd&&{}\cdot
|1-x|^{\frac23+\frac23i(\nu_a-2\nu_b+\nu_c+2\nu_d)}
\int \frac{d^2\rho_1\,d^2\rho_2}{|\rho_{12}|^4}
\onetworat acadcd-a 
\cdot{}
\cr\crd
&&{}\cdot
\onetworat abbdad-b \onetworat bccdbd-c 
\cdot{}\crd&&{}\cdot
\onetworat bdcdbc+d
\cr
&=&\cconst \frac{g^6}{16} 4 \frac2{(2\pi)^3} \xi(\nu_d)\,(4\nu_d^2+1)^2\;
\Xi(-\nu_a,-\nu_b,-\nu_c,\nu_d;x,x^*)\,,
&eq:psip3pbfkl}%
where the last equality defines the function $\Xi$.  Note that all the factors
in~(\ref{eq:psip3pbfkl}) are conformally invariant by themselves.  The function
$\Xi$ is defined as the integral multiplied with the powers of the anharmonic
ratios.  It was defined with the arguments $-\nu_a,\ldots,-\nu_c$.  Unlike a
general four-point function, our four-point function contains three complex
conjugated wave functions, so this notation facilitates comparison and
generalisation of our result.

The function~$\Xi$ is symmetric under simultaneous exchange of the Pomeron
coordinates and the $\tilde\nu\,$s ($\tilde\nu_{a/b/c}=-\nu_{a/b/c},
{}\tilde\nu_d=\nu_d$).  This follows from the fact that both the part of the
vertex without the $\Psi$ function (see (\ref{eq:our4point})) and the integral
of the BFKL term (\ref{eq:p3pbfkl}) have this symmetry.  Note that a
permutation of the coordinates changes the anharmonic ratio which is the last
argument of~$\Xi$.

\subsubsection{The $\alpha$ terms}

In this section we will calculate the $\Psi$ functions corresponding to the
terms containing the function~$\alpha$.  We will see that they can be written
in a form quite similar to~$\Psi_\bfklind$.

We will start with the term which arose from the set $M$ having one element,
$V^{(1)}$.  It contained a sub-integral with an expression which had to be
regularised.  The same integral has already occurred in a term of the $1\to2$
Pomeron vertex.  The associated integral operator is conformally invariant and
was found to have the eigenvalue~$1/2\,\xi(\nu)$~\cite{Lotter:1996vk},
\dmath2{
\int d^2\rho_1\,
\bigg[2\pi\,\delta^2(\rho_{13})\,\ln\epsilon
+\frac{|\rho_{23}|^2}{|\rho_{12}|^2|\rho_{13}|^2}
{}&&\theta\!\left(\frac{|\rho_{13}|}{|\rho_{12}|}-\epsilon\right)\bigg]
\,\cEz12a=
\cr&&=
\frac12\xi(\nu_a)\,\cEz23a \,,
&eq:bfkllike}%
and we have to use this relation under an integral over $\rho_2$ and 
$\rho_3$. Inserting it we get for this part of the vertex:
\dmath2{
V_{\pom\to3\pom}^{(1)}&=&
-\cconst \frac{g^6}{16} 4 \frac2{(2\pi)^3}(4\,\nu_d^2+1)^2
\sumabc\int \frac{d^2\rho_2\,d^2\rho_3}{|\rho_{23}|^4}
\cEz23b \cdot{}
\crd
&&\qquad
{}\cdot\cEz23c \eEz23d\,\frac12\xi(\nu_a)\,\cEz23a
\cr&=&
-\cconst \frac{g^6}{16} 4 \frac2{(2\pi)^3}
(\xi(\nu_a)+\xi(\nu_b)+\xi(\nu_c))\,(4\,\nu_d^2+1)^2
\int \frac{d^2\rho_2\,d^2\rho_3}{|\rho_{23}|^4}
\cdot{}&eq:p3p1xi\cr\crd&&\hskip -5mm{}\cdot
\cEz23a\cEz23b\cEz23c\eEz23d\,.
}%
Since the integrand except for the $\xi$ function is symmetric in the Pomeron
indices $a$, $b$ and $c$, we could remove the sum over the permutations and
insert a sum over $\xi$s instead.  The result closely resembles the BFKL
term~(\ref{eq:p3pbfkl}), except for the sign and the argument of the $\xi$
function.  Therefore we can immediately write down the $\Psi$ function of this
part of the vertex, 
\dmath2{
\Psi_{(1)}&{=}&
-\cconst \frac{g^6}{16} 4 \frac2{(2\pi)^3}
(\xi(\nu_a)+\xi(\nu_b)+\xi(\nu_c)) \,(4\nu_d^2+1)^2\;
\Xi(-\nu_a,-\nu_b,-\nu_c,\nu_d;x,x^*) \,.
\qquad
&eq:psip3p1}

The next group of $\alpha$ terms, $V^{(2,1)}$ contains a new integral.  We will
see later that this integral occurs in none of the other terms.  Therefore we
will just write down $\Psi_{(2,1)}$ without defining a function analogous 
to~$\Xi$.  Plugging (\ref{eq:p3p21}) into~(\ref{eq:getpsi}), we get:
\dmath2{
\Psi_{(2,1)}&=&
\cconst \frac{g^6}{16} 4 \frac2{(2\pi)^3}(4\,\nu_d^2+1)^2\;
|x|^{-\frac13+\frac23i(-2\nu_a+\nu_b+\nu_c-\nu_d)}\;
|1-x|^{\frac23+\frac23i(\nu_a-2\nu_b+\nu_c+2\nu_d)}
\cdot{}\cr&&{}\cdot
\psithree acadcd-a \psithree abbdad-b \psithree bccdbd-c \psithree bdcdbc+d
\cdot{}\cr&&{}\cdot
\sumabc \int d^2\rho_1\,d^2\rho_2
\frac1{|\rho_{12}|^4}
\cEz12a\eEz12d\cdot{}
\cr
&&{}\cdot
\int d^2\rho_3\,\left[2\pi\,\delta^2(\rho_{23})\ln\epsilon+
        \frac{|\rho_{12}|^2}{|\rho_{13}|^2|\rho_{23}|^2}
        \theta\!\left(\frac{|\rho_{23}|}{|\rho_{13}|}-\epsilon\right)\right]
\cEz13b
\cdot{}\cr&&{}\cdot
\cEz13c\,.
&eq:psip3p21}

There remain the integrals $V^{(2,2)}$ and $V^{(3)}$.  Looking at the formulas
(\ref{eq:p3p22}) and~(\ref{eq:p3p3}), one can see that they are very similar.
One has to be aware that the only distinction between different terms is the
number of Pomerons attached to a given pair of coordinates, (12), (13) or~(23).
Which of the three Pomerons $a$, $b$ and $c$ is attached to which does not
matter since all permutations are summed up.  Also, the indices of the
integration variables are immaterial because renaming them changes nothing.
Only one pair of indices is distinguished from the others by the fact that the
upper Pomeron with index $d$ is attached to it.

Taking all this into account, the only difference between $V^{(2,2)}$ and
$V^{(3)}$ is the following: In $V^{(2,2)}$, each of the three Pomerons attached
to the vertex from below has a different pair of coordinates as their
arguments.  In $V^{(3)}$, none of the lower Pomerons has the same pair of
coordinate arguments as the upper ($d$) Pomeron.  Instead, two of them have the
same pair of arguments.  (This can also be seen in the graphical representation
in Figure~\ref{fig:p3palpha} when one realises that the upper two lines have
the same coordinates as the outer lines below.)  This implies that $V^{(3)}$
can be obtained from $V^{(2,2)}$ by exchanging $\rho_d$ with the external
coordinate of one of the two lower Pomerons which have the same arguments
($\rho_b$ or $\rho_c$ in~(\ref{eq:p3p3})), and simultaneously exchanging
{\hbox{$\nu_d\leftrightarrow-\nu_b$}} or {\hbox{$\nu_d\leftrightarrow-\nu_c$}},
respectively.  Then the Pomeron $b$ resp.\ $c$ has the arguments which
previously none of the lower Pomerons had, and Pomeron $d$ has the same
arguments as one of the lower Pomerons.  This is exactly the situation in~$V^{(2,2)}$
(modulo permutation of the lower Pomerons and/or renaming of the integration
variables). We hence have the relation 
\dmath1{
V^{(3)}_{\pom\to3\pom}&
\putunder{$\longleftarrow\!\!\!-\!\!\!-\!\!\!-\!\!\!-\!\!\!-\!\!\!
-\!\!\!-\!\!\!-\!\!\!-\!\!\!-\!\!\!-\!\!\!-\!\!\!-\!\!\!\longrightarrow$}
{$\rho_d\leftrightarrow\rho_b, \nu_d\leftrightarrow-\nu_b$}
&V^{(2,2)}_{\pom\to3\pom} \,.
&eq:V3toV22\cr\noalign{\vskip2mm}}%
Therefore it can already be stated without any calculation that
$V^{(2,2)}$ and $V^{(3)}$ lead to the same type of integral.

In the following, we will calculate~$\Psi_{(3)}$.  $\Psi_{(2,2)}$ can then be
obtained by exchanging the coordinates and conformal dimension according
to~(\ref{eq:V3toV22}).  Inserting (\ref{eq:p3p3}) into the formula for $\Psi$, 
Eq.\ (\ref{eq:getpsi}), we get:
\dmath2{
\Psi_{(3)}&=&
-\cconst \frac{g^6}{16} 4 \frac2{(2\pi)^3} (4\,\nu_d^2+1)^2\;
|x|^{-\frac13+\frac23i(-2\nu_a+\nu_b+\nu_c-\nu_d)}\;
|1-x|^{\frac23+\frac23i(\nu_a-2\nu_b+\nu_c+2\nu_d)}
\cdot{}\crd&&{}\cdot
\psithree acadcd-a \psithree abbdad-b \psithree bccdbd-c \psithree bdcdbc+d
\cdot{}\crd&& {}\cdot
\sumabc\int \frac{d^2\rho_1\,d^2\rho_2\,d^2\rho_3\,}
                 {|\rho_{12}|^2|\rho_{13}|^2|\rho_{23}|^2}
\cEz12a\cEz13b
\cdot{}\crd&&{}\cdot
\cEz13c\eEz23d 
\cr&=&
-\cconst \frac{g^6}{16} 4 \frac2{(2\pi)^3} (4\,\nu_d^2+1)^2\;
\sumabc|x|^{-\frac13+\frac23i(-2\nu_a+\nu_b+\nu_c-\nu_d)}\;
\cdot{}\crd&&{}\cdot
|1-x|^{\frac23+\frac23i(\nu_a-2\nu_b+\nu_c+2\nu_d)}
\int \frac{d^2\rho_1\,d^2\rho_2\,d^2\rho_3\,}
          {|\rho_{12}|^2|\rho_{13}|^2|\rho_{23}|^2}
\onetworat acadcd-a 
\cdot{}
\crd
&&{}\cdot
\onethreerat abbdad-b \onethreerat bccdbd-c 
\cdot{}
\crd
&&{}\cdot
\twothreerat bdcdbc+d
\cr
&=&
-\cconst \frac{g^6}{16} 4 \frac2{(2\pi)^3} (4\,\nu_d^2+1)^2\;
\sumabc \Upsilon(-\nu_b,-\nu_c,-\nu_a,\nu_d;x,x^*)\,.
&eq:psip3p3}%
The step from the first to the second expression is not trivial.  It is possible 
only because the vertex without the $\Psi$ function is invariant under
permutations of the Pomerons $a$, $b$ and $c$, as can be seen
from~(\ref{eq:our4point}).  (In fact it is symmetric under all permutations of
conformal fields if one takes care of the different sign of~$\nu_d$.)  It is
this part of the vertex which we have written into the sum over the
permutations, even though the symmetry is not obvious any more.  The sum over
the permutations also extends to~$x$ in the sense that it is transformed into a
different anharmonic ratio by a permutation of the coordinates.  In fact, all
six anharmonic ratios occur in the sum, some with a minus sign.

The function $\Upsilon$ in (\ref{eq:psip3p3}) is defined as follows:
\dmath2{
\Upsilon(-&&\nu_b,-\nu_c,-\nu_a,\nu_d;x,x^*)=
|x|^{-\frac13+\frac23i(-2\nu_a+\nu_b+\nu_c-\nu_d)}\;
|1-x|^{\frac23+\frac23i(\nu_a-2\nu_b+\nu_c+2\nu_d)}
\cdot{}\crd&&{}\cdot
\int \frac{d^2\rho_1\,d^2\rho_2\,d^2\rho_3\,}
                                {|\rho_{12}|^2|\rho_{13}|^2|\rho_{23}|^2}
\onetworat acadcd-a \onethreerat abbdad-b
\cdot{}
\crd
&&\qquad\qquad\qquad{}\cdot
\onethreerat bccdbd-c \twothreerat bdcdbc+d\,.
&eq:Upsilon}%
We have defined it so that the conformal dimensions of the two Pomerons with
the same coordinate arguments come first.  It is symmetric in these first two
arguments provided the corresponding coordinates are exchanged as well, which
leads to a different anharmonic ratio, $1/x$ instead of~$x$, 
\dmath1{
\Upsilon(-\nu_b,-\nu_c,-\nu_a,\nu_d;x,x^*)&=&
\Upsilon(-\nu_c,-\nu_b,-\nu_a,\nu_d;\frac1x,\frac1{x^*}) \,.
&eq:Upssymm}

The conformally invariant function $\Psi_{(2,2)}$ corresponding to~$V^{(2,2)}$
can be obtained from $\Psi_{(3)}$ by way of~(\ref{eq:V3toV22}).  The only
slight difficulty is the anharmonic ratio~$x$.  Exchanging the coordinates
$\rho_d\leftrightarrow\rho_b$ also changes the anharmonic ratio on which
$\Upsilon$ still depends, 
\dmath2{
x=\crossrat abcdacbd \,
&\longrightarrow& \, 
\crossrat adbcacbd=1-x\,. 
&eq:xchange}%
Therefore we obtain for $\Psi_{(2,2)}$
\dmath2{
\Psi_{(2,2)}&=&
\cconst \frac{g^6}{16} 4 \frac2{(2\pi)^3} (4\,\nu_d^2+1)^2\;
\sumabc \Upsilon(-\nu_d,-\nu_c,-\nu_a,\nu_b;1-x,1-x^*)\,.
&eq:psip3p22\cr
}

\section{Summary and Conclusions}
\label{sumandconcl}

In this paper we have calculated the perturbative $1\to 3$ Pomeron 
vertex in the colour glass condensate. We have used the 
framework of the extended GLLA which is particularly 
well suited for the calculation of multi-Pomeron vertices. 
The $1\to 3$ Pomeron vertex is obtained by projecting the 
$2 \to 6$ reggeised gluon vertex onto BFKL Pomeron 
states. The resulting $1\to 3$ Pomeron vertex is local in 
rapidity. In the present paper we have taken into account only 
Pomeron states with conformal spin $n=0$ which are 
the leading states at high energies. It is in 
principle straightforward to extend our calculation to 
states with nonvanishing conformal spin. 

Let us now summarize the result of the calculation of the 
$1\to3$ Pomeron vertex performed in the previous section. 
We have found that the $1\to3$ Pomeron vertex has the 
form of a conformal four-point function, 
\dmath2{
V_{\pom\to3\pom}(\{\rho_i\};\{\nu_i\})&=&\Psi(x,x^*;\{\nu_i\})\prod_{i<j}
|\rho_{ij}|^{-\frac{2}{3} + 2i\left(-\tilde\nu_i-\tilde\nu_j
+\frac13\sum_k \tilde\nu_k\right)}
\,,
&eq:Vp3p4point}%
where the $\rho_i$ are the coordinates of the four Pomerons in 
impact parameter space, 
$x=\icrossrat abcdacbd$, and we have $\tilde\nu_i=-\nu_i$ 
for $i=a,b,c$ and $\tilde\nu_d=\nu_d$.

The freedom which remains in the otherwise fixed form of the vertex, the
function $\Psi$, can be written as a sum of five terms:
\dmath2{
\Psi(x,x^*;\{\nu_i\})&=&
\Psi_\bfklind(x,x^*;\{\nu_i\})+\Psi_{(1)}(x,x^*;\{\nu_i\})
\cr
&&\hskip -2mm{}+\Psi_{(2,1)}(x,x^*;\{\nu_i\})+\Psi_{(2,2)}(x,x^*;\{\nu_i\})
+\Psi_{(3)}(x,x^*;\{\nu_i\})\,.
&eq:psip3psum}%
These five terms have the following form:
\dmath{2.5}
{
\Psi_\bfklind&&(x,x^*;\{\nu_i\})=
\cconst \frac{g^6}{2(2\pi)^3} \xi(\nu_d)\,(4\nu_d^2+1)^2\;
\Xi(-\nu_a,-\nu_b,-\nu_c,\nu_d;x,x^*)
&eq:psip3pbfklsummary\cr
\Psi_{(1)}&&(x,x^*;\{\nu_i\})=
-\cconst \frac{g^6}{2(2\pi)^3}
(\xi(\nu_a)+\xi(\nu_b)+\xi(\nu_c)) \,(4\nu_d^2+1)^2\;
\cdot{}\crd&&\qquad\qquad\qquad\qquad\qquad{}\cdot
\Xi(-\nu_a,-\nu_b,-\nu_c,\nu_d;x,x^*)
&eq:psip3p1summary\cr
\Psi_{(2,1)}&&(x,x^*;\{\nu_i\})=
\cconst \frac{g^6}{2(2\pi)^3}(4\,\nu_d^2+1)^2\;
|x|^{-\frac13+\frac23i(-2\nu_a+\nu_b+\nu_c-\nu_d)}\;
\cdot{}\crd&&{}\cdot
|1-x|^{\frac23+\frac23i(\nu_a-2\nu_b+\nu_c+2\nu_d)}
\psithree acadcd-a \psithree abbdad-b 
\cdot{}\crd&&{}\cdot
\psithree bccdbd-c \psithree bdcdbc+d
\sumabc \int d^2\rho_1\,d^2\rho_2
\frac1{|\rho_{12}|^4}
\cdot{}
\crd
&&{}\cdot
\cEz12a\eEz12d
\int d^2\rho_3\,\bigg[2\pi\,\delta^2(\rho_{23})\ln\epsilon
\crd&&{}
+        \frac{|\rho_{12}|^2}{|\rho_{13}|^2|\rho_{23}|^2}
        \theta\!\left(\frac{|\rho_{23}|}{|\rho_{13}|}-\epsilon\right)\bigg]
\cEz13b\cEz13c
&eq:psip3p21summary\cr
\Psi_{(2,2)}&&(x,x^*;\{\nu_i\})=
\cconst \! \frac{g^6}{2(2\pi)^3} (4\,\nu_d^2+1)^2
\hskip -2mm \sumabc \hskip -2mm\Upsilon(-\nu_d,-\nu_c,-\nu_a,\nu_b;1{-}x,1{-}x^*)
&eq:psip3p22summary\cr
\Psi_{(3)}&&(x,x^*;\{\nu_i\})=
-\cconst \frac{g^6}{2(2\pi)^3} (4\,\nu_d^2+1)^2\;
\sumabc \Upsilon(-\nu_b,-\nu_c,-\nu_a,\nu_d;x,x^*) \,.
&eq:psip3p3summary}%
We recall that the colour constant $\cconst$ has for general~$N_c$ 
the form 
\dmath1{
\cconst=\frac1{N_c}(N_c^2-4)(N_c^2-1)\,,
&&&eq:constC}%
and the $\xi(\nu)$ is given in (\ref{eq:xi}). 
The sums in $\Psi_{(2,1)}$, $\Psi_{(2,2)}$ and~$\Psi_{(3)}$ run over all six
simultaneous permutations of the conformal dimensions $\nu_{a/b/c}$ and the
corresponding coordinates $\rho_{a/b/c}$.  That entails that $x$ is
replaced by a different cross ratio according to the permutation of the
$\rho\,$s.  The two integral expressions $\Xi$ and~$\Upsilon$ in the 
expressions above are defined as 
\dmath2{
\Xi&&(-\nu_a,-\nu_b,-\nu_c,\nu_d;x,x^*)=
|x|^{-\frac13+\frac23i(-2\nu_a+\nu_b+\nu_c-\nu_d)}\;
|1-x|^{\frac23+\frac23i(\nu_a-2\nu_b+\nu_c+2\nu_d)}
\cdot{}\crd&&\qquad\qquad{}\cdot
\int \frac{d^2\rho_1\,d^2\rho_2}{|\rho_{12}|^4}
\onetworat acadcd-a \onetworat abbdad-b
\cdot{}
\crd
&&\qquad\qquad\qquad\qquad{}\cdot
\onetworat bccdbd-c \onetworat bdcdbc+d
&eq:xisummary\cr
\Upsilon&&(-\nu_b,-\nu_c,-\nu_a,\nu_d;x,x^*)=
|x|^{-\frac13+\frac23i(-2\nu_a+\nu_b+\nu_c-\nu_d)}\;
|1-x|^{\frac23+\frac23i(\nu_a-2\nu_b+\nu_c+2\nu_d)}
\cdot{}\crd&&{}\cdot
\int \frac{d^2\rho_1\,d^2\rho_2\,d^2\rho_3\,}
                                {|\rho_{12}|^2|\rho_{13}|^2|\rho_{23}|^2}
\onetworat acadcd-a \onethreerat abbdad-b
\cdot{}
\crd
&&\qquad\qquad\qquad\quad{}\cdot
\onethreerat bccdbd-c \twothreerat bdcdbc+d
\,.
&eq:upssummary}%

Note that all terms of the $1 \to 3$ Pomeron vertex calculated here 
are of the same order in powers of $N_c$. This is in contrast to the 
$1 \to 2$ Pomeron vertex in which there are two contributions 
one of which is suppressed by two powers of $N_c$ with respect 
to the other. 

We should point out again that we have only calculated the 
contribution to the $1 \to 3$ Pomeron vertex coming from the 
$2 \to 6$ gluon vertex $V_{2 \to 6}$ obtained in \cite{Bartels:1999aw}. 
The situation here is analogous to the case of the $1 \to 2$ Pomeron 
vertex which has been calculated in \cite{Bartels:1994jj,Lotter:1996vk} 
from the $2 \to 4$ gluon vertex $V_{2 \to 4}$. The additional 
contribution to that vertex coming from the first term (the reggeising term) 
in (\ref{d4solutionpics}) upon projection onto Pomeron eigenstates 
as described in section \ref{sec:pvertices} could be quite important 
but has not yet been studied in detail. 
Similarly, we expect that there are also other contributions 
to the $1 \to 3$ Pomeron 
vertex coming from the reggeising terms in the six-gluon amplitude 
which have been described briefly in section \ref{subsec:eglla}. 
These additional contributions 
clearly need to be computed in order to obtain a full picture of 
the vertex. For the part of the six-gluon amplitude which is a 
superposition of four-gluon amplitudes this amounts to 
performing a projection of the vertices $I,J$ and $L$ obtained in 
\cite{Bartels:1999aw}. Technically, that calculation is similar 
to the one we have presented here. 
First steps in this direction have been performed in \cite{Volkerthesis}. 

The $1 \to 3$ Pomeron vertex gives an additional contribution to 
the $t$-channel evolution of hadronic scattering processes at high 
energy. It would of course be very important to determine the relative 
importance of this vertex with respect to the usual fan diagrams in which 
the transition from one to three Pomerons occurs only via the iteration 
of the $1 \to 2$ Pomeron vertex. 
In the expansion in $N_c$ the direct $1 \to 3$ Pomeron vertex is 
clearly suppressed with respect to the transition 
via the iterated $1 \to 2$ Pomeron vertex, i.\,e.\ with 
respect to the corresponding fan diagram. But in order to determine 
the relative importance of the $1 \to 3$ Pomeron vertex one also has to 
consider its numerical value which, if found to be large, can potentially 
compensate the suppression due to factors of $N_c$. 
In this context one also has to consider the additional contributions to the 
$1 \to 3$ Pomeron vertex coming from the reggeising terms in the 
six-gluon amplitude, which we have mentioned above and 
which have not yet been calculated. 
These contributions are not necessarily of the same order in the 
$N_c$ expansion, in particular they can be less suppressed with respect 
to the iterated $1 \to 2$ Pomeron vertex in the fan diagrams. 
These issues need to be studied in detail before the question of the 
phenomenological relevance of the $1 \to 3$ Pomeron vertex can be answered.  

As already pointed out it would be important to calculate the numerical 
value of the perturbative $1 \to 3$ Pomeron vertex at least for the 
leading Pomeron states with $h=1/2$. The integral expressions 
occurring in the vertex are in principle similar to those 
in the $1 \to 2$ Pomeron vertex, which have been computed in 
\cite{Korchemsky:1997fy}. The problem of computing the value 
of the $1 \to 3$ Pomeron vertex, although clearly more complicated, 
can probably be approached with the same techniques. 

The $2 \to 4$ and the $2 \to 6$ gluon vertex 
in the EGLLA have a very similar structure which is most 
conveniently expressed in terms of the function $G$. Based 
on this structure it has been conjectured that also higher $2 \to 2n$ 
gluon vertices exist for arbitrary $n$ \cite{Ewerz:2001uq}. 
According to their conjectured momentum structure 
it appears likely that they also give rise to higher $1 \to n$ 
Pomeron vertices which are local in rapidity. 
But we have seen above that also the colour structure of the 
$2 \to 6$ gluon vertex was crucial for identifying the terms 
contributing to the $1 \to 3$ Pomeron vertex. 
In order to establish the existence of general $1 \to n$ 
Pomeron vertices one would therefore not only have to prove 
the conjectured momentum structure. In addition it would 
be necessary to determine also the explicit 
colour tensors of the $2 \to 2n$ gluon vertices. 

An important conceptual implication of our result is that it 
indicates an inequivalence of the EGLLA and 
the dipole picture of high energy scattering occurring 
when one goes beyond the approximation of fan diagrams 
involving only the $1 \to 2$ Pomeron vertex. 
We have found that in the EGLLA a $1\to3$ Pomeron vertex 
exists which is local in rapidity, whereas in the dipole picture 
such a local vertex is absent according to \cite{Braun:1997nu}. 
We also find that our formula for the $1 \to 3$ Pomeron vertex 
does not coincide with the form conjectured for this vertex in 
\cite{Peschanski:1997yx} based on a generalization of the 
$1 \to 2$ Pomeron vertex in the dipole picture. 
Nevertheless, it appears possible that the 
two are related to each other in suitable limits or via duality 
or bootstrap relations. Also here the other contributions from 
the reggeising parts of the six-gluon amplitude to the $1\to3$ 
Pomeron vertex might be relevant. 
It will of course also be very interesting to study whether the $1 \to 3$ 
Pomeron vertex found in the present paper can be related to the 
analogous vertex obtained in the approach due to Balitsky 
in \cite{Balitsky:2001mr}. The answer to these questions will be 
very valuable for understanding the relations and characteristic 
differences between different approaches to the physics of the 
colour glass condensate. 

\section*{Acknowledgements}

We would like to thank Jochen Bartels, Gregory Korchemsky, and 
Robi Peschanski for helpful discussions and Thomas Bittig for 
comments on the manuscript. 

\appendix

\section*{Appendix}
\label{app:dE}
\subsection*{Derivatives of the conformal eigenfunctions \boldmath$E^{(\nu,n)}$}

When proving the conformal transformation properties of the $G$~function we
have used the explicit form of derivatives of the conformal eigenfunction
$E^{(\nu,n)}$. These derivatives will be calculated in this appendix.

The starting point is the explicit form of the conformal eigenfunction 
$E^{(\nu,n)}$ of the BFKL kernel in  position space as given 
in (\ref{eq:Enun_app}). 
$\rho_a$ in that formula is the external coordinate of the Pomeron state. 

The differential operators which occur in the $G$~function are:
$\Delta_1\Delta_2$, $\nabla_1\Delta_2$ and a single~$\Delta_1$.  Expressed as
derivatives with respect to the complex coordinates, they become:
$16\,\del_1\del^*_1\del_2\del^*_2$, $8\,\del^*_1\del_2\del^*_2$ and
$4\,\del_1\del^*_1$, respectively. 
We can completely separate the conjugated from the
unconjugated coordinates.

We first calculate the gradient with respect to one coordinate of the 
eigenfunction~$E^{(\nu,n)}$. 
\dmath2{
\nabla_1 E^{(\nu,n)}(\rho_{1},\rho_{2})
&=&2\,\del^*_1 E^{(\nu,n)}(\rho_{1},\rho_{2})
=2 \left(\frac{\rho_{12}}{\rho_{1a}\rho_{2a}}\right)^{\frac{1+n}2+i\nu}
\del^*_1
\left(\frac{\rho_{12}^*}{\rho_{1a}^*\rho_{2a}^*}\right)^{\frac{1-n}2+i\nu}
\cr
&=&2 \left(\frac{\rho_{12}}{\rho_{1a}\rho_{2a}}\right)^{\frac{1+n}2+i\nu}
\left(\frac{1-n}2+i\nu\right)
\left(\frac{\rho_{12}^*}{\rho_{1a}^*\rho_{2a}^*}\right)^{-\frac{1-n}2+i\nu}
\frac1{{\rho^*_{1a}}^2}
\cr
&=&(1-n+2\,i\nu)\frac{\rho^*_{2a}}{\rho^*_{1a}\rho^*_{12}}
E^{(\nu,n)}(\rho_{1},\rho_{2}) \,.
&eq:nablE_app}%
Performing the differentiation with respect to the non-conjugated coordinate 
yields an analogous result, and similar results hold for the other coordinate, 
\dmath2{
\nabla^*_1 E^{(\nu,n)}(\rho_{1},\rho_{2})&=&
(1+n+2\,i\nu)\frac{\rho_{2a}}{\rho_{1a}\rho_{12}}
E^{(\nu,n)}(\rho_{1},\rho_{2})
&eq:nablE1star_app\cr
\nabla_2 E^{(\nu,n)}(\rho_{1},\rho_{2})&=&
-\,(1-n+2\,i\nu)\frac{\rho^*_{1a}}{\rho^*_{2a}\rho^*_{12}}
E^{(\nu,n)}(\rho_{1},\rho_{2})
&eq:nablE2_app\cr
\nabla^*_2 E^{(\nu,n)}(\rho_{1},\rho_{2})&=&
-\,(1+n+2\,i\nu)\frac{\rho_{1a}}{\rho_{2a}\rho_{12}}
E^{(\nu,n)}(\rho_{1},\rho_{2})
\,.
&eq:nablE2star_app}%
The derivatives with respect to~$\rho_2$ get a minus sign because the result is
expressed in~$\rho_{12}$ instead of~$\rho_{21}$.

Having derived the gradient, we can immediately give the
Laplacians of a Pomeron wave function~$E^{(\nu,n)}$. Since the differentiations
with respect to (un)conjugated variables add only factors of variables of the
same type, the two differentiations contained in a Laplacian are 
independent.  We obtain
\dmath2{
\Delta_1 E^{(\nu,n)}(\rho_{1},\rho_{2})&=&
\big((1+2\,i\nu)^2-n^2\big)\frac{|\rho_{2a}|^2}{|\rho_{1a}|^2|\rho_{12}|^2}
E^{(\nu,n)}(\rho_{1},\rho_{2})
&eq:laplE1_app\cr
\Delta_2 E^{(\nu,n)}(\rho_{1},\rho_{2})&=&
\big((1+2\,i\nu)^2-n^2\big)\frac{|\rho_{1a}|^2}{|\rho_{2a}|^2|\rho_{12}|^2}
E^{(\nu,n)}(\rho_{1},\rho_{2})\,.
&eq:laplE2_app}

The combination of a gradient and a Laplacian requires some more work.  Here
the gradient affects also the additional coordinate moduli
in~(\ref{eq:laplE2_app}).  In this case we find 
\dmath2{
\nabla_1\Delta_2 E^{(\nu,n)}(\rho_{1},\rho_{2})
&=&\nabla_1 \big((1+2\,i\nu)^2-n^2\big)
        \frac{|\rho_{1a}|^2}{|\rho_{2a}|^2|\rho_{12}|^2}
        E^{(\nu,n)}(\rho_{1},\rho_{2})
\cr
&=&\big((1+2\,i\nu)^2-n^2\big)\bigg(
2\left(\del^*_1\frac{|\rho_{1a}|^2}{|\rho_{2a}|^2|\rho_{12}|^2}\right)
\cr&&{}
+\frac{|\rho_{1a}|^2}{|\rho_{2a}|^2|\rho_{12}|^2}
(1-n+2\,i\nu)\frac{\rho^*_{2a}}{\rho^*_{1a}\rho^*_{12}}\bigg)
        E^{(\nu,n)}(\rho_{1},\rho_{2})\,.
&eq:gradlaplderive}%
The derivative of the first factor becomes
\be
\del^*_1\frac{|\rho_{1a}|^2}{|\rho_{2a}|^2|\rho_{12}|^2}
=\frac{\rho_{1a}}{|\rho_{2a}|^2\rho_{12}}
\del^*_1\frac{\rho^*_{1a}}{\rho^*_{12}}
=\frac{|\rho_{1a}|^2}{|\rho_{2a}|^2|\rho_{12}|^2}
        \frac{-\rho^*_{2a}}{\rho^*_{1a}\rho^*_{12}}\,.
\ee
This term can easily be added to the derivative of the second factor, giving
the constant prefactor $(-1-n+2\,i\nu)$.  After factorising the prefactor of
(\ref{eq:laplE2_app}) and combining it with the new factor, we obtain the
result:
\dmath2{
&&\nabla_1\Delta_2 E^{(\nu,n)}(\rho_{1},\rho_{2})={}
\cr&&\qquad
-\big(4\,\nu^2+(n+1)^2\big)(1-n+2\,i\nu)
\frac{\rho^*_{2a}}{\rho^*_{1a}\rho^*_{12}}
\frac{|\rho_{1a}|^2}{|\rho_{2a}|^2|\rho_{12}|^2}
E^{(\nu,n)}(\rho_{1},\rho_{2}).
&eq:gradlaplresult}

The additional differentiation necessary to apply a double Laplacian to%
~$E^{(\nu,n)}$ is again independent of the one just performed and can be 
done analogously. The moduli of $\rho_{1a}$ and~$\rho_{2a}$ then cancel out. 
The result is
\begin{equation}
\label{eq:LapLapE}
\Delta_1\Delta_2 E^{(\nu,n)}(\rho_{1},\rho_{2})
=\big(4\,\nu^2+(n+1)^2\big)\big(4\,\nu^2+(n-1)^2\big)
 \frac1{|\rho_{12}|^4} E^{(\nu,n)}(\rho_{1},\rho_{2})\,.
\end{equation}


\begin{thebibliography}{99}
\addcontentsline{toc}{part}{Bibliography}

\bibitem{McLerran:1993ka}
L.~D.~McLerran and R.~Venugopalan,
Phys.\ Rev.\ D {\bf 49} (1994) 3352
[arXiv:hep-ph/9311205].

\bibitem{McLerran:1993ni}
L.~D.~McLerran and R.~Venugopalan,
Phys.\ Rev.\ D {\bf 49} (1994) 2233
[arXiv:hep-ph/9309289].

\bibitem{McLerran:1994vd}
L.~D.~McLerran and R.~Venugopalan,
Phys.\ Rev.\ D {\bf 50} (1994) 2225
[arXiv:hep-ph/9402335].

\bibitem{Jalilian-Marian:1996xn}
J.~Jalilian-Marian, A.~Kovner, L.~D.~McLerran and H.~Weigert,
Phys.\ Rev.\ D {\bf 55} (1997) 5414
[arXiv:hep-ph/9606337].

\bibitem{Iancu:2000hn}
E.~Iancu, A.~Leonidov and L.~D.~McLerran,
Nucl.\ Phys.\ A {\bf 692} (2001) 583
[arXiv:hep-ph/0011241].

\bibitem{Ferreiro:2001qy}
E.~Ferreiro, E.~Iancu, A.~Leonidov and L.~McLerran,
Nucl.\ Phys.\ A {\bf 703} (2002) 489
[arXiv:hep-ph/0109115].

\bibitem{Iancu:2003xm}
E.~Iancu and R.~Venugopalan,
arXiv:hep-ph/0303204.

\bibitem{Balitsky:1995ub}
I.~Balitsky,
Nucl.\ Phys.\ B {\bf 463} (1996) 99
[arXiv:hep-ph/9509348].

\bibitem{Balitsky:1998kc}
I.~Balitsky,
Phys.\ Rev.\ Lett.\  {\bf 81} (1998) 2024
[arXiv:hep-ph/9807434].

\bibitem{Balitsky:1998ya}
I.~Balitsky,
Phys.\ Rev.\ D {\bf 60} (1999) 014020
[arXiv:hep-ph/9812311].

\bibitem{Balitsky:2001re}
I.~Balitsky,
Phys.\ Lett.\ B {\bf 518} (2001) 235
[arXiv:hep-ph/0105334].

\bibitem{Balitsky:2001mr}
I.~I.~Balitsky and A.~V.~Belitsky,
Nucl.\ Phys.\ B {\bf 629} (2002) 290
[arXiv:hep-ph/0110158].

\bibitem{Blaizot:2002xy}
J.~P.~Blaizot, E.~Iancu and H.~Weigert,
Nucl.\ Phys.\ A {\bf 713} (2003) 441
[arXiv:hep-ph/0206279].

\bibitem{Kovchegov:1999yj}
Y.~V.~Kovchegov,
Phys.\ Rev.\ D {\bf 60} (1999) 034008
[arXiv:hep-ph/9901281].

\bibitem{Mueller:1993rr}
A.~H.~Mueller,
Nucl.\ Phys.\ B {\bf 415} (1994) 373.

\bibitem{Mueller:1994jq}
A.~H.~Mueller and B.~Patel,
Nucl.\ Phys.\ B {\bf 425} (1994) 471
[arXiv:hep-ph/9403256].

\bibitem{Mueller:gb}
A.~H.~Mueller,
Nucl.\ Phys.\ B {\bf 437} (1995) 107
[arXiv:hep-ph/9408245].

\bibitem{Chen:1995pa}
Z.~Chen and A.~H.~Mueller,
Nucl.\ Phys.\ B {\bf 451} (1995) 579.

\bibitem{Kuraev:fs}
E.~A.~Kuraev, L.~N.~Lipatov and V.~S.~Fadin,
Sov.\ Phys.\ JETP {\bf 45} (1977) 199
[Zh.\ Eksp.\ Teor.\ Fiz.\  {\bf 72} (1977) 377].

\bibitem{Balitsky:ic}
I.~I.~Balitsky and L.~N.~Lipatov,
Sov.\ J.\ Nucl.\ Phys.\  {\bf 28} (1978) 822
[Yad.\ Fiz.\  {\bf 28} (1978) 1597].

\bibitem{Bartels:1980pe}
J.~Bartels,
Nucl.\ Phys.\ B {\bf 175} (1980) 365.

\bibitem{Kwiecinski:1980wb}
J.~Kwieci{\'n}ski and M.~Prasza{\l}owicz,
Phys.\ Lett.\ B {\bf 94} (1980) 413.

\bibitem{Bartels:1978fc}
J.~Bartels,
Nucl.\ Phys.\ B {\bf 151} (1979) 293.

\bibitem{Bartels:unp}
J.~Bartels,
DESY 91-074 (unpublished) 

\bibitem{Bartels:1992ym}
J.~Bartels,
Phys.\ Lett.\ B {\bf 298} (1993) 204.

\bibitem{Bartels:1993ih}
J.~Bartels,
Z.\ Phys.\ C {\bf 60} (1993) 471.
 
\bibitem{Bartels:1994jj}
J.~Bartels and M.~W\"usthoff,
Z.\ Phys.\ C {\bf 66} (1995) 157.

\bibitem{Bartels:1995kf}
J.~Bartels, L.~N.~Lipatov and M.~W\"usthoff,
Nucl.\ Phys.\ B {\bf 464} (1996) 298
[arXiv:hep-ph/9509303].

\bibitem{Lotter:1996vk}
H.~Lotter,
Ph.\,D.\ Thesis, Hamburg University 1996, DESY 96-262,
arXiv:hep-ph/9705288.
 
\bibitem{Braun:1997nu}
M.~A.~Braun and G.~P.~Vacca,
Eur.\ Phys.\ J.\ C {\bf 6} (1999) 147
[arXiv:hep-ph/9711486].

\bibitem{Bartels:1999aw}
J.~Bartels and C.~Ewerz,
JHEP {\bf 9909} (1999) 026
[arXiv:hep-ph/9908454].
 
\bibitem{Ewerz:1999gk}
C.~Ewerz,
Phys.\ Lett.\ B {\bf 472} (2000) 135
[arXiv:hep-ph/9911225].

\bibitem{Ewerz:2001fb}
C.~Ewerz,
JHEP {\bf 0104} (2001) 031
[arXiv:hep-ph/0103260].

\bibitem{Ewerz:2001uq}
C.~Ewerz,
Phys.\ Lett.\ B {\bf 512} (2001) 239
[arXiv:hep-ph/0105181].

\bibitem{Volkerthesis}
V.\ Schatz,
Ph.\,D.\ Thesis, Heidelberg University 2003, HD-THEP-03-21, 
arXiv:hep-ph/0307326.

\bibitem{Ferreiro:2002kv}
E.~Ferreiro, E.~Iancu, K.~Itakura and L.~McLerran,
Nucl.\ Phys.\ A {\bf 710} (2002) 373
[arXiv:hep-ph/0206241].

\bibitem{Kramer:tr}
A.~Kr\"amer and H.~G.~Dosch,
Phys.\ Lett.\ B {\bf 252} (1990) 669.
 
\bibitem{Kraemer:rc}
A.~Kr\"amer and H.~G.~Dosch,
Phys.\ Lett.\ B {\bf 272} (1991) 114.

\bibitem{Dosch:1992cu}
H.~G.~Dosch, E.~Ferreira and A.~Kr\"amer,
Phys.\ Lett.\ B {\bf 289} (1992) 153.
 
\bibitem{Dosch:1994ym}
H.~G.~Dosch, E.~Ferreira and A.~Kr\"amer,
Phys.\ Rev.\ D {\bf 50}, 1992 (1994)
[arXiv:hep-ph/9405237].

\bibitem{Dosch:1987sk}
H.~G.~Dosch,
Phys.\ Lett.\ B {\bf 190} (1987) 177.
 
\bibitem{Dosch:ha}
H.~G.~Dosch and Y.~A.~Simonov,
Phys.\ Lett.\ B {\bf 205} (1988) 339.
 
\bibitem{Simonov:1987rn}
Y.~A.~Simonov,
Nucl.\ Phys.\ B {\bf 307} (1988) 512.

\bibitem{Nachtmann:1991ua}
O.~Nachtmann,
Annals Phys.\  {\bf 209} (1991) 436.

\bibitem{Donnachie:en}
A.~Donnachie, G.~Dosch, O.~Nachtmann and P.~Landshoff, 
``Pomeron Physics And QCD,''
{\it  Cambridge University Press, Cambridge 2002}. 

\bibitem{Korchemsky:1997fy}
G.~P.~Korchemsky,
Nucl.\ Phys.\ B {\bf 550} (1999) 397
[arXiv:hep-ph/9711277].

\bibitem{Bialas:1997xp}
A.~Bialas, H.~Navelet and R.~Peschanski,
Phys.\ Lett.\ B {\bf 427} (1998) 147
[arXiv:hep-ph/9711236].

\bibitem{Bialas:1997ig}
A.~Bialas, H.~Navelet and R.~Peschanski,
Phys.\ Rev.\ D {\bf 57} (1998) 6585
[arXiv:hep-ph/9711442].

\bibitem{Ewerz:2003xi}
C.~Ewerz,
arXiv:hep-ph/0306137.

\bibitem{Kovchegov:2003dm}
Y.~V.~Kovchegov, L.~Szymanowski and S.~Wallon,
arXiv:hep-ph/0309281.

\bibitem{Peschanski:1997yx}
R.~Peschanski,
Phys.\ Lett.\ B {\bf 409} (1997) 491
[arXiv:hep-ph/9704342].

\bibitem{Lipatov:1985uk}
L.~N.~Lipatov,
Sov.\ Phys.\ JETP {\bf 63} (1986) 904
[Zh.\ Eksp.\ Teor.\ Fiz.\  {\bf 90} (1986) 1536].

\bibitem{Bartels:2002au}
J.~Bartels, M.~G.~Ryskin and G.~P.~Vacca,
Eur.\ Phys.\ J.\ C {\bf 27} (2003) 101
[arXiv:hep-ph/0207173].

\end{thebibliography}
\end{document}